\documentclass[a4paper,fleqn]{cas-sc}

\usepackage[numbers]{natbib}
\usepackage{subfig}
\usepackage{algorithm}
\usepackage{algpseudocode}

\def\tsc#1{\csdef{#1}{\textsc{\lowercase{#1}}\xspace}}
\tsc{WGM}
\tsc{QE}

\begin{document}
\let\WriteBookmarks\relax
\def\floatpagepagefraction{1}
\def\textpagefraction{.001}

\shorttitle{DADIN}    

\shortauthors{<short author list for running head>}  

\title [mode = title]{DADIN: Domain Adversarial Deep Interest Network for Cross Domain Recommender Systems}  



%

\author[1]{Menglin Kong}[type=editor,style=chinese]





\credit{Conceptualization, Methodology, Software, Validation, Investigation, Writing - original draft, Writing - review \& editing, Visualization.}

\affiliation[1]{organization={School of Mathematics and Statistics},
            addressline={Central South University}, 
            city={Changsha},
            postcode={410083}, 
            state={Hunan},
            country={China}}

\author[1]{Muzhou Hou}[type=editor,style=chinese]




\credit{Conceptualization, Methodology, Resources Writing - review \& editing, Supervision, Project administration, Funding acquisition.}

\author[2]{Shaojie Zhao}[type=editor,style=chinese]


\credit{Software, Validation, Formalan alysis, Data curation, Writing - review \& editing, Visualization.}

\affiliation[2]{organization={School of Mathematics, Physics and Statistics},
            addressline={Shanghai University of Engineering Science}, 
            city={Shanghai},
            postcode={201620}, 
            state={Shanghai},
            country={China}}

\author[3]{Feng Liu}[type=editor,style=chinese]


\credit{Conceptualization, Methodology, Software, Validation, Investigation, Supervision, Resources}

\affiliation[3]{organization={School of Mathematics and Statistics},
	addressline={The University of Melbourne}, 
	city={Parkville},
	postcode={3010},
	state={Melbourne},
	country={Australia}}
        
\author[1]{Ri Su}[type=editor,style=chinese]

\credit{Writing - review \& editing.Supervision, Project administration.}

\author[1,4]{Yinghao Chen}[orcid=0000-0002-9176-8394,type=editor,style=chinese]

\cormark[1]

\ead{chenyinghao@csu.edu.cn}

\credit{Conceptualization, Methodology, Software, Validation, Investigation, Writing-original draft, Writing-review \& editing, Visualization, Supervision, Project administration Funding acquisition.}

\cortext[1]{Corresponding author}

\affiliation[4]{organization={Eastern Institute for Advanced Study},
	addressline={Yongriver Institute of Technology}, 
	city={Ningbo},
	postcode={315201},
	state={Zhejiang},
	country={China}}



\begin{abstract}
Click-Through Rate (CTR) prediction is one of the main tasks of the recommendation system, which is conducted by a user for different items to give the recommendation results. Cross-domain CTR prediction models have been proposed to overcome problems of data sparsity, long tail distribution of user-item interactions, and cold start of items or users. In order to make knowledge transfer from source domain to target domain more smoothly, an innovative deep learning cross-domain CTR prediction model, Domain Adversarial Deep Interest Network (DADIN) is proposed to convert the cross-domain recommendation task into a domain adaptation problem. The joint distribution alignment of two domains is innovatively realized by introducing domain agnostic layers and specially designed loss, and optimized together with CTR prediction loss in a way of adversarial training. It is found that the Area Under Curve (AUC) of DADIN is 0.08\% higher than the most competitive baseline on Huawei dataset and is 0.71\% higher than its competitors on Amazon dataset, achieving the state-of-the-art results on the basis of the evaluation of this model performance on two real datasets. The ablation study shows that by introducing adversarial method, this model has respectively led to the AUC improvements of 2.34\% on Huawei dataset and 16.67\% on Amazon dataset.
\end{abstract}



\begin{keywords}
Cross-domain Recommendation \sep Click-through Rate \sep Cold-Start Recommender Systems \sep Transfer Learning
\end{keywords}

\maketitle

\section{Introduction}\label{1}
\begin{figure}[ht]
	\centering
	\includegraphics[width=0.45\textwidth]{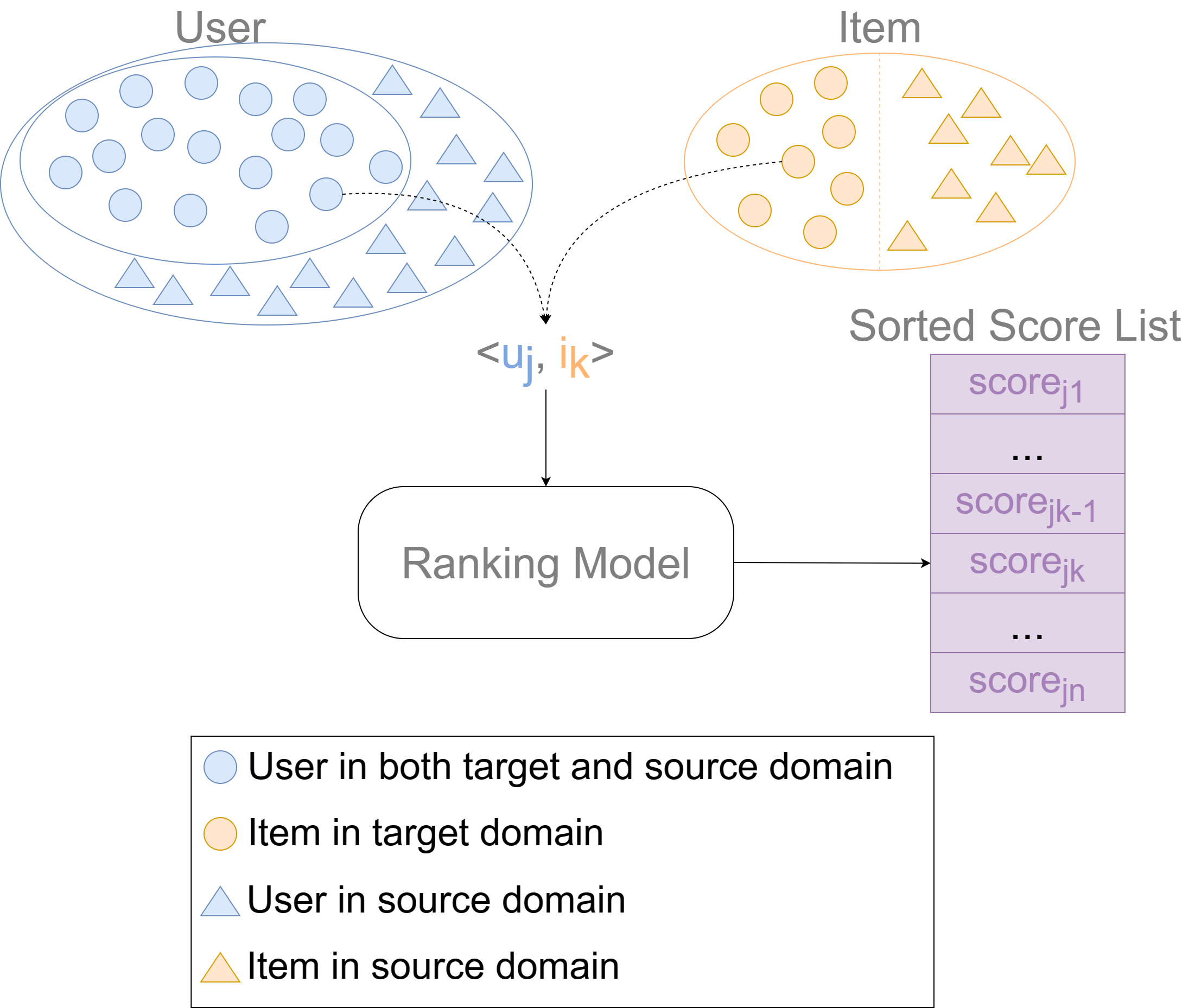}
	\caption{{\bf Understanding CTR prediction task from a ranking perspective}}\label{fig:rs}
\end{figure}

For Click-Through Rate (CTR) prediction, it is usually necessary to learn the features representation of user $u_k$ and item $i_j$, so that they can be projected into the same feature space to better describe the similarity between users and items \citep{singh2008relational}. Through the calculation the similarity of feature embeddings between user $u_k$ and item $i_j$ in the same feature space, the score (predicted CTR) $score_{jk}$ for the user-item pair $<u_j, i_k>$ is obtained. However, when learning feature space mapping, problems such as data sparsity \citep{pan2010transfer,kang2019semi,song2021coarse}, long tail distribution of user-item interactions \citep{zhang2021model}, and cold start of items or users \citep{kang2019semi} are usually encountered. In recent years, researchers have proposed to use cross-domain CTR models to solve these problems. The original single-domain perspective is changed because the knowledge of other domains is adopted to help complete the task of the target domain. Problems from the single-domain perspective, such as cold start and data sparsity, can also be solved by transferring cross-domain knowledge \citep{khan2017cross}. The general cross-domain recommendation scenario is illustrated in Figure \ref{fig:rs}. User set in target domain is usually the subset of that in source domain, and item set is not shared in the two domains.

The existing methods to achieve cross-domain CTR prediction mainly focus on: 1. Self-supervised or semi-supervised pre-training explicit mapping methods, such as EMCDR \citep{man2017cross}, TMCDR \citep{zhu2021transfer}, PTUPCDR \citep{zhu2022personalized}, etc.; 2. CTR models with the twin-towers structure that perform implicit mapping based on Domain Adaptation (DA), such as CoNet \citep{hu2018conet}, CATN \citep{zhao2020catn}, MiNet \citep{ouyang2020minet}, etc. It is generally believed that compared with the two-stage training model (pre-training explicit mapping methods), the end-to-end model (implicit mapping methods) can better optimize the objective function and add prior knowledge to the model through adding regularizations on the model parameters. Therefore, a cross-domain CTR model is designed based on DA with the twin tower structure. The infrastructure of this model looks at the architecture of MiNet \citep{ouyang2020minet} but considers organizing and processing model inputs in differently.

Meanwhile, the classic theory H$\Delta$ h-Divergence \citep{ben2010theory}, proposed by Ben-David, has a generalized error boundary for DA theory on classic binary classification tasks. It is found that reducing the classification loss in the source domain can help reduce the classification loss in the target domain. Yaroslav prove the effectiveness of the Domain Adversarial method based on this theory and propose the Domain Adversarial Neural Network (DANN) \citep{ganin2016domain}. As the subsets of the binary task of CTR prediction, both the DA method and the Domain Adversarial method are with strict theoretical proof. Therefore, the domain adversarial method is introduced the domain adversarial method as the guiding principle to align the source and target domain distribution. Guided by the Domain Adversarial approach, Domain Agnostic Layer is added into this model and two kinds of special domain confusion loss are introduced as prior knowledge to align both the marginal and conditional distribution of the source and target domain.

\begin{figure}[ht]
	\centering
	\includegraphics[width=0.45\textwidth]{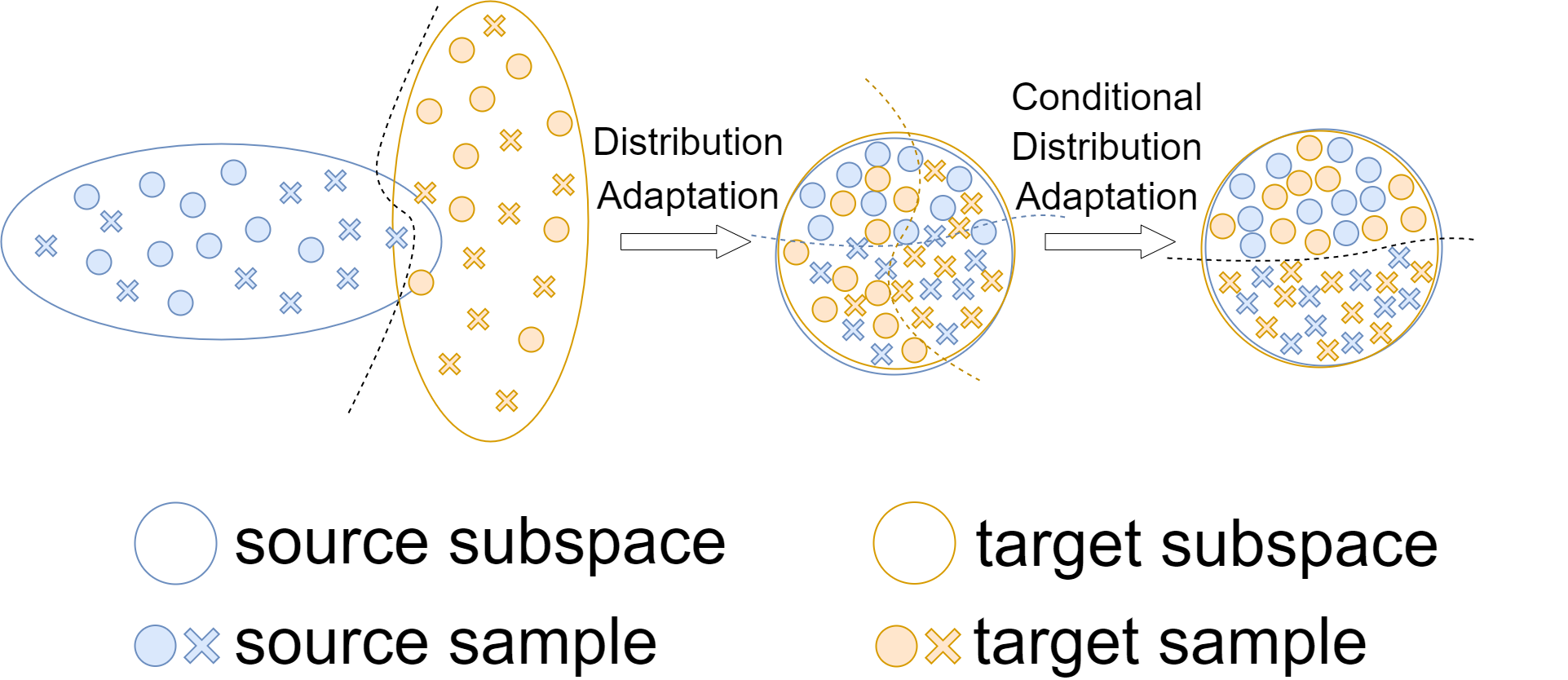}
	\caption{{\bf Achievement of the marginal and conditional distribution alignment of target and source domain samples.}}\label{fig:cda}
\end{figure}

In the real applications of DA, the source domain data is usually more affluent than the target domain data in terms of classes, which brings a problem: those classes that only exist in the source domain will have a negative transfer on the adaptation results. Long et al. point out that this problem is called partial transfer learning, that is, samples from the source domain can be transferred on condition that they are related to the target domain. Selective Adversarial Networks (SAN) \citep{cao2018partial} is proposed to deal with the partial transfer problems. In order to introduce sample-level constraints, several domain discriminators related to those classes whose inputs are weighted by labels prediction are adopted by SAN. Thus, samples in the source domain that are more similar to those in the target domain are transferred. This phenomenon may exist in our cross-domain CTR task (binary classification task). When the distribution alignment is carried out, the marginal distribution of the source domain and target domain may be aligned in the joint feature space. However their conditional distributions are not similar, as shown in the middle part of Figure \ref{fig:cda}. In addition, existing domain adaptation methods are usually implemented by using two classifiers for two domains, and the classification task in the target domain might be handicapped by the introduction of source domain information. In DADIN, the joint distribution alignment is achieved by introducing skip-connection-based domain agnostic layer and domain confusion loss method \citep{yu2019transfer} that focus on aligning the marginal distribution and conditional distribution of data from the target domain and source domain at the same time. The same classifier is used for the joint training of source and target domains, and the domain discriminator is added to provide domain confusion loss. The effect of DA of this method is shown in the right part of Figure \ref{fig:cda}.

The main contributions of this method are summarized as follows:

\begin{itemize}
	\item The DADIN method is proposed by using Domain Adversarial approach to enhance the knowledge transfer from the source domain to the target domain. The joint distribution alignment is achieved by introducing a skip-connection-based domain agnostic layer and domain confusion loss method. The domain adversarial method is firstly used on cross-domain CTR prediction models.
	\item To achieve more fine-grained interest extraction, DADIN uses an item-level attention mechanism to strengthen the feature representation of the user's historical behavior and interest-level attention to perform weighted concatenation of embeddings from different fields.
	\item The performance of DADIN is evaluated based on two large real datasets and compared with the current optimal single-domain and cross-domain CTR models, achieving state-of-the-art (SOTA) results.
	\item The validity of the method based on Domain Adversarial Learning and all components of the model is verified through rigorously designed experiments on artificial datasets and ablation studies on real datasets.
\end{itemize}

The rest of this paper is organized as follows: Section \ref{RELATED WORK} introduces the related works of this paper, focusing on the single-domain and cross-domain CTR models and deep network domain adaption. Then, Section \ref{METHOD} describes the proposed model DADIN and introduces the global and intra-class domain confusion loss. Next, Section \ref{Experiments} provides conceptual experiment results to verify the effectiveness of the domain adversarial method and the comparison results of this model with single-domain and cross-domain baseline models in two real datasets. Finally, Section \ref{Conclusion} is the conclusion and prospects.

\section{Related works}\label{RELATED WORK}

\subsection{Cross-domain Recommendation}

CTR prediction is the core task in the application of recommendation system. Based on the similarity between user preference vector and item feature vector, situations such as collaborative filtering and list-wise recommendation need to be completed according to the estimated click-through rate. However, in most recommendation situations based on implicit feedback, it is difficult to fully mine the user's personal preferences because the user's behavior cannot be fully observed (i.e. lack of real negative samples) \citep{zhang2016deep, qu2016product, shan2016deep}, which also leads to the inconsistency of distribution between the training set used by the model and inference space. The algorithm based on cross-domain recommendation system is a way to solve this kind of problem. Through the introduction of the behavior information of users in the target domain from the perspective of other domains, the robustness of the training model is enhanced, so that it has the ability of out-of-distribution (OOD) generalization. Cross-domain recommendation methods can be broadly classified into three categories: \textbf{1)} collaborative \citep{singh2008relational,hu2018conet}, \textbf{2)} content-based \citep{elkahky2015multi,zhang2016collaborative}, and \textbf{3)} hybrid \citep{lian2017cccfnet} methods.
The collaborative method is inspired by the cross-stick network (CSN) \citep{misra2016cross} in the CV field. CSN extracts the features of the images in the two fields independently, and achieves the goal of bidirectional knowledge transfer by linear combination of the feature maps in the high-dimensional space. Collaborative cross networks (CoNet) \citep{hu2018conet} is the representative of the application of collaborative methods in the field of cross-domain recommendation systems. By introducing cross connection units, the linear combination is changed into linear transformation, which realizes more fine-grained and sparse knowledge transfer. However, collaborative methods pay more attention to the mutual promotion of classification tasks in the two fields, and the shared transfer matrix may lead to negative transfer. The content-based method maps the information from the user side and the item side into a common embedding space through a shared embedding layer, then explicitly or implicitly models their interaction effects. Multi-view Deep Learning model (MV-DNN) \citep{elkahky2015multi} is an extension of content-based method in cross-domain recommendation. MVDNN regards user/item information from multiple domains as their expression in multiple views, and uses multiple mapping layers to map their features independently into high-dimensional feature space for infomation fusion. Literature \citep{zhang2016collaborative} introducing multi-modal item-side information on the basis of MVDNN is conducive to achieve more comprehensive and fine-grained feature extraction. However, an independent DNN for each view feature needs to be introduced in content-based methods, resulting in complex model structure and parameter redundancy. The hybrid method is the combination of the above two. In \citep{lian2017cccfnet} a neural network with a double-tower structure is proposed. The two-layers Multilayer Perceptron (MLP) is responsible for extraction of the user and item vectors based on matrix factorization (MF) algorithms for collaborative filtering, while the DNN is responsible for encoding the rich user and item side features. The Mixed Interest Network (MiNet) jointly models three types of user interest in a hierarchical attention-based manner. Our method refers to the hierarchical attentions of MiNet when modeling the user's historical behavior, but introducing the domain agnostic sub-network based on deep adversarial learning to enhance the representation of cross-domain features.

\subsection{Deep Domain Adaptation}

DA is the core concept in Transfer Learning (TL), which generally means that mapping the samples from the source and target domains to the same feature space is conducive to the similar distribution of samples in the source and target domains of the feature space \citep{ghifary2014domain,krizhevsky2017imagenet,tzeng2014deep,long2015learning}. The classification model trained on its source domain samples can be directly transferred to the target domain to solve the problem of insufficient training samples. DaNN \citep{ghifary2014domain} is an early domain adaptation method based on deep learning, which is composed of feature mapping layer and prediction layer. The Maximum Mean Discrepancy (MMD) measure of source and target domain samples is added to the loss function as a regularization term to achieve the alignment of source domain and target domain in the feature space.
Differently, in Ganin \citep{ganin2016domain} draw on the Min-Max game idea of Generative Adversarial Nets (GAN), compares the module of feature mapping to the generator in GAN, and introduce the domain discriminator to identify representation from different domains. When the domain discriminator cannot complete the task, it means that the extracted features no longer contain domain information, which is an implicit domain adaptation process. Recently, the concept of dynamic distribution adaptation in traditional TL is further extended to deep adversarial networks \citep{zhao2022learning}, proving that there is also a mismatch between marginal and conditional distribution in adversarial networks.
The cross-domain recommendation model draws on \citep{yu2019transfer} 's idea of adaptively aligning the distribution of the source and target domain. By introducing both global and intra-class domain confusion loss into the loss function, it is found the marginal distribution alignment and conditional distribution alignment of source and target domain data in the same embedding space. Furthermore, CTR predictors applicable to the two domains can be learned in the same embedding space by using both source and target domain data, and the joint training of domain discrimination and CTR prediction can be realized by referring to the Gradient Reverse Layer (GRL) proposed in \citep{ganin2016domain}.

\section{Method}\label{METHOD}
We introduce our method from the following aspects: problem formulation and symbol description, model overview, loss function and model parameter learning.

\subsection{Overview}

\subsubsection{Problem Formulation}
In the cross-domain CTR prediction task, there are two item sets, namely the item set $I_t$ of the target domain (e.g., advertisements) and the item set $I_s$of the source domain (e.g., news) and a common user set $U$. In the datasets, there are interactions $y_{kj}^{t}$ between user $u_k $ and the item  $i_{j}^{t}$ from target domain and interactions $y_{kj}^{s}$ between the user $u_k $ and the item  $i_{j}^{s}$ from source domain. In real business situations, the size of user-item interactions from the target domain that the model can learn is much smaller than that from the source domain. The users involved in the interaction records in the target domain are a subset of the users in the source domain, and there is a relation of inclusion between the two. It can be assumed that user $u_k$ has a domain-agnostic consistent tendency towards browsing preferences for items from different domains, so the learning of the CTR prediction model of the target domain with fewer interactive records can be prompted by using the source domain samples through domain adaption methods. In the cold start situation, when the interaction record related to the new user $u_k$ does not exist in the target domain but exists in the source domain, the knowledge of the source domain can enhance the CTR prediction of the target domain.

\begin{figure*}[ht]
	\centering
	\includegraphics[width=\textwidth]{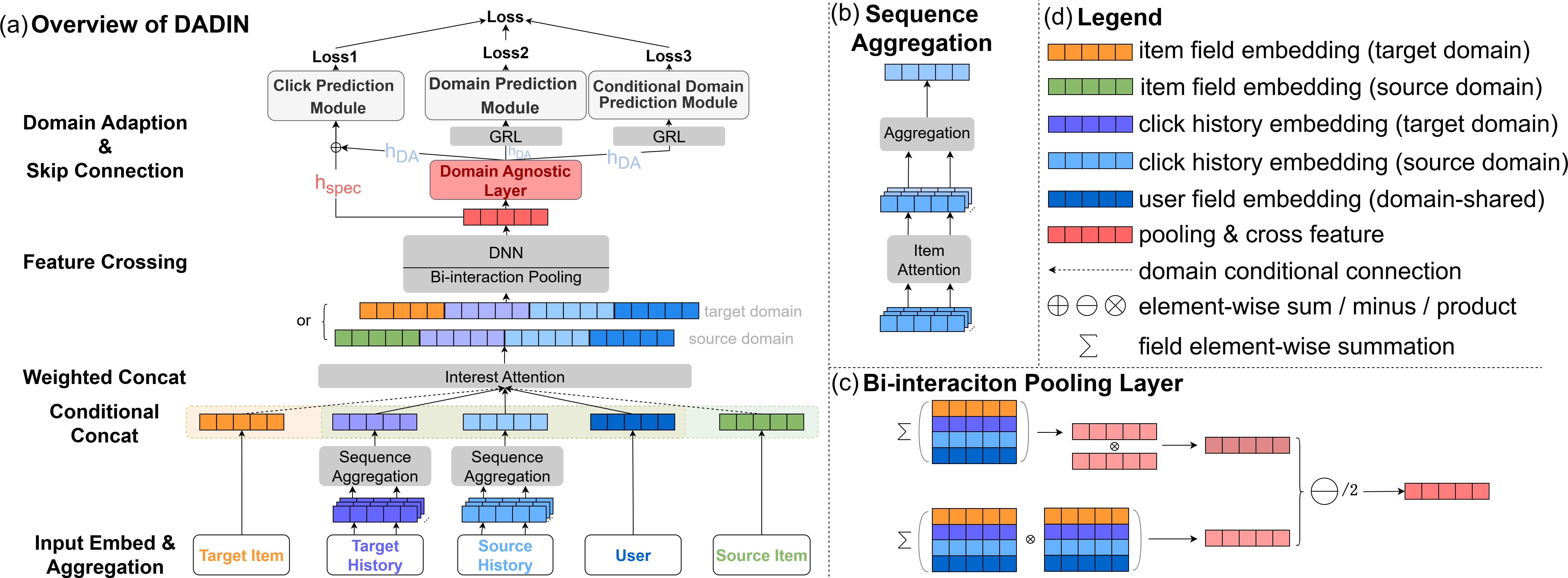}
	\caption{{\bf Structure of Domain Adversarial Deep Interest Network (DADIN). (a) is an overview of DADIN. (b) and (c) illustrate two special operations, that is, \textit{Sequence Aggregation} and \textit{Bi-interaction Pooling}. (d) is an explanation of those graphics that appear previously.}}\label{fig:structure}
\end{figure*}

\subsubsection{Forward Progress}

The Domain Adversarial Deep Interest Network (DADIN) is proposed in Figure \ref{fig:structure}. Figure \ref{fig:structure} (a) inputs from both domains are first split into five parts, features from target item field $\textbf{v}_i^{t}$, features from user field $\textbf{v}_i^{u}$ (which consist of features from user profile $\textbf{v}_i^{up}$, target history field $\textbf{v}_i^{th}$ and source history field $\textbf{v}_i^{sh}$) as well as features from source item field (one instance has features from one domain, either target domain or source domain), they are projected to a dense but low-dimensional space by the \textit{Embedding Layer} (corresponding to Input Embed \& Aggregation in Figure \ref{fig:structure} (a)). The DADIN contains a \textit{Single-domain Item  Attention Layer} and \textit{Sequence Aggregation} operation to reinforce the embedding of user history behavior feature from the target and source domain, respectively. After that, a \textit{Interest Attention Layer} is used to achieve weighted concat of embeddings from the user field and target domain item field (or source domain item field, there is only one situation in an instance). Then the concatenated embeddings are feedforward to the feature crossing module, which consists of \textit{Bi-interaction Pooling Layer} to perform explicit feature crossing and a \textit{DNN} to performs implicit feature crossing. Different from regular cross domain recommendation models, the DADIN uses a \textit{Domain Agnostic Layer} to extract domain-agnostic features $h_{DA}$ from both domains, then perform element-wise summation on the input and output of the \textit{Domain Agnostic Layer}, which is a skip layer connection to aggregate domain-agnostic information $h_{DA}$ and domain-specific information $h_{spec}$. In this way, the DADIN uses the source domain information to enhance the feature representation of the target domain. A common \textit{Label Predictor} for both domains takes the final feature representation $h = h_{DA} + h_{spec}$ to give the CTR predictions. A \textit{Domain Classifier} and two \textit{Conditional Domain Classifiers} take the domain-agnostic information $h_{DA}$ to calculate global domain confusion loss $L_d^{\dagger}$ and intra-class domain confusion loss $L_d^0, L_d^1$ respectively.

\subsection{Interest Extraction}

\subsubsection{Feature Embedding}
In the model, one instance (a record of the user's interaction with the item) has features from either the target domain item field or the source domain item field. For each instance, its features is divided into five parts; the same operations are conducted on features from the user profile field as features from the target domain item field and source domain item field. For a user $u_k$, it is supposed that he has featured "user\_ID=100009, age=5, residence=13, city=191", there are have four embedding vectors $x_{uid=100009}$, $x_{age=5}$, $x_{city=191}$, $x_{residence=13} \in \mathbb{R}^d$, then field pooling is performed through element-wise summation on the four vectors to obtain the aggregated embeddings $p_k \in \mathbb{R}^{d}$ for $u_k$, that is,
\begin{equation}
	p_k =x_{uid=100009} +x_{age=5}+x_{city=191}+x_{residence=13}.
\end{equation}

In different datasets, a user may have a different number of attributes, so its embedding for the user profile field is the aggregation of a different number of embedding vectors $\mathbf{x}_i \in \mathbb{R}^{d}$. In this way, embeddings from target domain item field $q_j^{t} \in \mathbb{R}^{d}$ and source domain item field $q_j^{s} \in \mathbb{R}^{d}$ can be obtained.
The embedding $hist_k$ of the corresponding recently clicked items is also obtained by embedding the clicked items in the sequence. In this process, different pooling methods are used, which will be covered this in detail in the next part.

\subsubsection{Sequence Aggregation}

When dealing with the user's historical behavior in the target domain and source domain, DADIN uses item attention and sequence aggregation. Denote the set of embedding vectors of recently clicked items and the aggregated representation as $\left \{ r_{tj} \right \} _k$, and $a_j$ respectively.
\begin{equation}
	\mathbf{a}_{j}=\sum_{t} \beta_{t} \mathbf{r}_{t j}, \beta_{t}=\frac{\exp \left(\tilde{\beta}_{t}\right)}{\sum_{t^{\prime}} \exp \left(\tilde{\beta}_{t^{\prime}}\right)},
	\label{eq:ag}
\end{equation}
\begin{equation}
	\tilde{\beta}_{t}=\mathbf{h}_{j}^{T} \operatorname{ReLU}\left(\mathbf{W}_{j}\left[\mathbf{r}_{t j} \|\mathbf{q}_{j} \| \mathbf{p}_{k} \| \mathbf{r}_{t j} \odot \mathbf{q}_{j}\right]\right),
\end{equation} 
where $\mathbf{W}_{j} \in \mathbb{R}^{d \times 4d}$ and $\mathbf{h}_{t} \in \mathbb{R}^d$ are learnable parameters. The notation "$\|$" is the vector concatenation operation. In Equation (\ref{eq:ag}), $r_{tj}\in \mathbb{R}^d$ reflects the embedding vector of clicked items $i_t^{t}$ in the target domain, $q_j\in \mathbb{R}^d$ reflects the embedding vector of candidate item $i_j^{t}$ in the target domain, $p_k\in \mathbb{R}^d$ reflects the embedding vector of user $u_k$, $\mathbf{r}_{tj} \odot \mathbf{q}_{j}\in \mathbb{R}^d$ reflects the interaction between history item $i_t^{t}$ and candidate item $i_j^{t}$from the target domain. Moreover, $\odot$ is the element-wise product operator. After the item-level attention and \textit{Sequence Aggregation} operation, the feature from target history field $\left \{r_{tj} \right \} _k\in \mathbb{R}^{d\times seqlen} $ ($seqlen$ is a hyperparameter, refers to the max length of clicked items of user $u_k$) is converted to a compact feature representation $\mathbf{a}_j^{k} \in \mathbb{R^d}$. Through the same processes, a compact feature representation $\mathbf{b}_j^{k} \in \mathbb{R^d}$ for a user $u_k$ and a candidate item $i_j^{(s)}$ from the source domain through.

\subsubsection{Interest Attention}
For one instance from the target domain, the vector $\mathbf{q}_j^{(t)} \in \mathbb{R}^{d}$ is from the target domain item field, the vector $\mathbf{p_k} \in \mathbb{R}^{d}$ is from user profile field, the vector $\mathbf{b}_j^{k} \in \mathbb{R^d}$ is from source history field, and the vector $\mathbf{a}_j^{k} \in \mathbb{R^d}$ is from target history field. It is natural to consider that those four types of embedding vectors can be concatenated to form a more informative feature representation $\mathbf{m} \in \mathbb{R}^{4 d}$ for the next layer, that is,
\begin{equation}
	\mathbf{m} \triangleq \left[\mathbf{q}_j^{(t)}\left\|\mathbf{p}_{k}\right\| \mathbf{a}_{j}^k \| \mathbf{b}_{j}^k\right].
	\label{eq:without_interest_attention}
\end{equation}

However, this combination of features introduces the assumption that three different types of information from the user field are of equal importance to the candidate item embedding vector $\mathbf{q}_j^{t} $, restricting the representation capacity of the model. Therefore, weighted concatenation is conducted for these four types of embeddings using \textit{Interest Attention}. Specifically, weights are assigned to embeddings $p_k,b_j^{k},a_j^{k}$ based on their correlation with embedding $\mathbf{q}_j^{t}$. A weighted concatenate feature representation $\mathbf{m}_{jk}$ is finally gotten:
\begin{equation}
	\mathbf{m}_{jk} \triangleq \left[\mathbf{q}_{j}^{t}\left\|v_{u} \mathbf{p}_{k}\right\| v_{t} \mathbf{a}_{j}^{k} \| v_{s} \mathbf{b}_{j}^{k}\right].
	\label{eq:interest_attention}
\end{equation}

These weights are computed as follows:
\begin{equation}
	\begin{array}{l}
		v_{u}=\exp \left(\mathbf{g}_{u}^{T} \operatorname{ReLU}\left(\mathbf{V}_{u}\left[\mathbf{q}_{j}^{t}\left\|\mathbf{p}_{k}\right\| \mathbf{a}_{j}^{k} \| \mathbf{b}_{j}^{k}\right]\right)+b_{u}\right), \\
		v_{s}=\exp \left(\mathbf{g}_{s}^{T} \operatorname{ReLU}\left(\mathbf{V}_{s}\left[\mathbf{q}_{j}^{t}\left\|\mathbf{p}_{k}\right\| \mathbf{a}_{j}^{k} \| \mathbf{b}_{j}^{k}\right]\right)+b_{s}\right), \\
		v_{t}=\exp \left(\mathbf{g}_{t}^{T} \operatorname{ReLU}\left(\mathbf{V}_{t}\left[\mathbf{q}_{j}^{t}\left\|\mathbf{p}_{k}\right\| \mathbf{a}_{j}^{k} \| \mathbf{b}_{j}^{k}\right]\right)+b_{t}\right),
	\end{array}
	\label{eq:cal_weight}
\end{equation}
where $\mathbf{V}_{*} \in \mathbb{R}^{d \times4d}$ is a matrix parameter, $\mathbf{g}_{*} \in \mathbb{R}^{d}$ is a vector parameter and $b_*$ is a scalar parameter. In Equation (\ref{eq:interest_attention}), it is found that these dynamic weights $v_u,v_t,v_s$ are calculated using all available information related to user $u_k$ and candidate item $i^{t}_j$  given all three types of user interests (user profile features, user history behaviors in source domain, user history behaviors in target domain), so these weights can measure the importance of embeddings reflecting user's interest from three different perspectives to the embedding of candidate item $i^{t}_j$. In the experiment, it is found that the value of dynamic weights obtained only through matrix multiplication and ReLU activation function is very small, which may lead to the weighted concatenate vector $\mathbf{m}_{jk}$ being dominated by $\mathbf{q}_j^{t}$. Therefore, an exponential function is acted on the results in Equation (\ref{eq:cal_weight}), mapping it to $\mathbb{R}^+$.

\subsubsection{Feature Crossing Layer}

The weighted concatenate feature representation $m_{jk}$ obtained in the above process only represents the features of each field; in the model design of the recommendation system, great importance is often attached to the feature crossing. Therefore, in consideration of reducing computational complexity and model parameters as much as possible, the \textit{Bi-Interaction Pooling Layer} \citep{zhang2016collaborative} is introduced to carry out explicit second-order crossing of features on the four field embeddings:
\begin{equation}
	f_{BI}(\mathbf{m}_{jk}; C_\mathbf{m}) = \sum^4_{i=1} \sum^4_{j=i+1} \mathbf{m}_{jk}^{i} c_i \odot \mathbf{m}_{jk}^{j} c_j,
\end{equation}
where $f_{BI}(\cdot)$ represents the Bi-interaction Pooling Layer and $C_m$ is its parameter, $\mathbf{m}_{jk}^{i}$ represents the embedding vector of the $i$ field of the weighted concatenate feature representation $\mathbf{m}_{jk}$. More intuitively, this operation is equal the following:
\begin{equation}
	f_{BI}(\mathbf{m}_{jk}; C_\mathbf{m}) = \frac{1}{2} \left | (\sum^4_{i=1} \mathbf{m}_{jk}^{(i)} c_i )^2 - \sum^4_{i=1} (\mathbf{m}_{jk}^{(i)} c_i)^2 \right |.
\end{equation}

Figure \ref{fig:structure} (c) gives a visual explanation of Bi-interaction Pooling. Through the Bi-interaction Pooling Layer, the feature representation of the instance changes from $\mathbf{m}_{jk} \in \mathbb{R}^{4d}$ to $\textbf{z} \in \mathbb{R}^{d}$. 

Naturally, the results obtained from $f_{BI}(\cdot)$, namely $\textbf{z} \in \mathbb{R}^{d}$ are feedforward into the DNN module containing two fully-connected (FC) layers for further implicit feature crossing. The output of the DNN $\textbf{h}_{spec} \in \mathbb{R}^{d}$ is obtained, which represents the combination of features from different fields and carries more fine-grained feature crossing signals.

\subsection{Domain Adaptation}

\subsubsection{Domain Agnostic Layer}

After the feature crossing module, the combined feature is gotten after field pooling and more fine-grained feature crossing in the feature space $\textbf{h}_{spec} \in \mathbb{R}^{d}$. However,the input to the model is organized by adding the domain label  $y_d$ for each input instance, which resulted in each instance containing only the features from the target or the source domain item field. That is to say, the combined feature $\textbf{h}_{spec} \in \mathbb{R}^{d}$ here is sampled from the data distribution $D_*$ of its domain. Considering that the model uses a common CTR predictor when outputting the CTR prediction $\hat{y}$, if the data distribution $D_s$ and $D_t$ of the two domains are far from each other, the direct transfer of knowledge from the source domain to the target domain will have an adverse effect on the results of the model, which is "negative transfer" phenomenon. Negative transfer can correspond to the state before conditional distribution alignment in Figure \ref{fig:cda}. Therefore, it is considered that the domain adversarial method tends to be used to realize the data distribution adaptation of the target and source domain.

Specifically, $\textbf{h}$ denotes the intermediate state before prediction modules, and it is composed of domain-agnostic information $\textbf{h}_{DA}$ and domain-related information $\textbf{h}_{spec}$. Their relationship can be described as $\textbf{h}$ is equals to $\textbf{h}=\textbf{h}_{DA}+\textbf{h}_{spec}$. \textit{Domain Agnostic Layer} is added to the model, which takes the  $\textbf{h}_{spec} \in \mathbb{R}^{d}$ as input and outputs the feature of the domain-agnostic part $\textbf{h}_{DA} \in \mathbb{R}^{d}$. $\textbf{h}_{DA} \in \mathbb{R}^{d}$ is wanted to confuse its domain discriminator into not knowing which domain the instance is from. Such inductive bias is introduced by domain confusion loss calculated with domain labels and outputs of the global and the intra-class domain discriminator. Finally, $\textbf{h}= \textbf{h}_{DA}+\textbf{h}_{spec}$ is gained through the skip connection.

\subsubsection{Label predictor \& Domain classifier}

In this section, the feature representation after the skip connection is first feedforward to the common label predictor of the target and the source domain to obtain the CTR prediction, namely $G(\textbf{h})$= $\hat{y}$. Specifically, our label predictor contains two FC layers and adds a Dropout layer between the two FC layers to alleviate the overfitting of the model. Formally, the label predictor is defined as follows:
\begin{equation}
	\textbf{h}_1=\text{Dropout}(W_1\textbf{h}+\textbf{b}_1),
\end{equation}
\begin{equation}
	\hat{y}= \sigma(W_2\textbf{h}_1+\textbf{b}_2),
	\label{eq:predict}
\end{equation}
where $W_1 \in \mathbb{R}^{100\times d}$, $\mathbf{\textbf{b}_1} \in \mathbb{R}^{100}$, $W_2\in \mathbb{R}^{1\times 100}$ and $\mathbf{b_2}\in \mathbb{R} $ are learnable parameters of the two FC layers, $\mathbf{h}_1\in \mathbb{R}^{100}$ is the intermediate state, $\text{Dropout}(\cdot)$ and $\sigma(\cdot)$ are the Dropout layer and Sigmoid function respectively.

On the other hand, the domain-agnostic feature representation $\mathbf{h}_{DA}$ into domain classifiers is feedforward to calculate the domain prediction $\hat{d}$. Because $\mathbf{h}_{DA}$ is wanted to deceive the domain classifiers, domain prediction $\hat{d}$ of the instance should be pretty different from the true domain label (domain confusion loss should be large). To realize the simultaneous optimization of CTR prediction loss, and domain confusion loss and the joint training of the model, the GRL is introduced.

GRL's mathematical formulation is defined as Rev(x), and its forward and backpropagation are:
\begin{equation}
	\begin{aligned}
		\text{Rev}(\mathbf{x})=\mathbf{x}, \\
		\frac{\mathrm{d} \text{Rev}}{\mathrm{d} \mathbf{x}} =-I,
	\end{aligned}
\end{equation}
where $I$ is an identity matrix. With GRL, the objective of minimizing CTR prediction loss can be realized on both domains and maximizing domain confusion simultaneously during the training stage. The details will be illustrated in Section \ref{loss}.

Concerning the domain classifiers, global domain classifiers $D^{\dagger}(\cdot)$ is obtained, which takes $\mathbf{h}_{DA}$ through GRL as input and outputs the domain prediction $\hat{d}$. The structure of $D^{\dagger}(\cdot)$ is,
\begin{equation}
	\hat{d}=\sigma(W_3\text{Rev}(\mathbf{h}_{DA})+b_3),
	\label{eq:domain}
\end{equation}
where $W_3\in \mathbb{R}^{1\times d}$, $b_3\in \mathbb{R}$ are learnable parameters of the global domain classifiers $D^{\dagger}$.

Two intra-class domain classifiers $D^0$ and $D^1$ are also innovatively introduced our model to alleviate partial transfer proposed in \citep{cao2018partial}; they have the same structure as $D^{\dagger}$. The optimization in $D^0$ and $D^1$ represents conditional distribution alignment in Figure \ref{fig:cda}. Different from $D^{\dagger}$, $D^1$ only takes $h_{DA}$ through GRL, which is predicted to be from class 1. Especially, a threshold $T$ is set, if  $\hat{y} $ from label predictor outnumbers $T$, then the instance is believed to come from class 1 and feedforward to $D^1$ to calculate domain prediction $\hat{d}^1$. For $D^0$, the situation is precisely the same.
\begin{equation}
	\hat{d}^1=\sigma(W_4\text{Rev}(\hat{y}\times \mathbf{h}_{DA})+\mathbf{h}_4),
	\label{eq:domain1}
\end{equation}
\begin{equation}
	\hat{d}^0=\sigma(W_5\text{Rev}((1-\hat{y})\times \mathbf{h}_{DA})+\mathbf{h}_5),
	\label{eq:domain0}
\end{equation}
By introducing the global domain classifier $D^{\dagger}$, the marginal distribution alignment of the target and source domain, and conditional distribution alignment are realized by using two intra-class domain classifiers $D^0$ and $D^1$. The DA manner based on the joint distribution alignment of features from both domains greatly enhances knowledge transfer from the source domain to the target domain show the competitive results of the method will be shown in Section \ref{Experiments}.

\subsection{Optimization} \label{loss}

The cross-entropy loss is adopted as the loss function. For the CTR prediction loss $L_y$ (\textit{Loss1} in Figure \ref{fig:structure} (a)), instances from both target and source domains are used to calculate:
\begin{equation}
	L_y = \frac{1}{n}\sum_{i=1}^{n} [y_i\log_{}{\hat{y}_i} +(1-y_i)\log_{}({1-\hat{y}_i})],
\end{equation}
where $y_i$ is the true label of the instance $i$, and $\hat{y}_i$ is the CTR prediction of the instance $i$ output by label predictor $G$, $n$ is the number of instances from both domains in the training set.

For the global domain confusion loss $L_d^{\dagger}$ (\textit{Loss2} in Figure \ref{fig:structure} (a)), the same instances is used to calculate the loss, but instances from target domain are assigned with domain label $d_i$ = 1, and those from source domain with domain label $d_i$ = 0, then there is:
\begin{equation}
	L_d^{\dagger} = \frac{1}{n}\sum_{i=1}^{n} [d_i\log_{}{\hat{d}_i} +(1-d_i)\log_{}({1-\hat{d}_i})],
\end{equation}
where $d_i$ is the domain label of the instance $i$, and $\hat{d}_i$ is the domain prediction of the instance $i$ output by global domain classifier $D^{\dagger}$.

The intra-class domain confusion loss $L_d^{1}$ and $L_d^{0}$ are introduced. Different from $L_y$ and $L_d^{\dagger}$, both $L_d^{1}$ and $L_d^{0}$ only incorporate a subset of the total instances to calculate the loss. Specifically, a threshold $T$ is set, if $\hat{y}_i\ge T $, then instance $i$ will be assigned to be the positive example regardless of its domain, the instance's feature representation $\mathbf{h}_{DA}$ will be feedforward to domain classifier $D^1$, and intra-class domain confusion loss $L_d^{1}$ can be calculated as:
\begin{equation}
	L_d^{1} = \frac{1}{\rho}\sum_{i=1}^{\rho} [d_i\log_{}{\hat{d}_i^1} +(1-d_i)\log_{}({1-\hat{d}_i^1})].
\end{equation}

On the contrary, when $\hat{y}_i< T$, the situation is exactly the same:
\begin{equation}
	L_d^{0} = \frac{1}{\tau}\sum_{i=1}^{\tau} [d_i\log_{}{\hat{d}_i^0} +(1-d_i)\log_{}({1-\hat{d}_i^0})],
\end{equation}
where $\rho$ denotes the number of instances whose CTR prediction $y_i$ outnumbers $T$ and $\tau$ is the number of instances whose CTR prediction $y_i$ is less than $T$, $n=\rho+\tau$, $\hat{d}_i^1$ and $\hat{d}_i^0$ are the domain prediction of intra-class domain classifiers $D^1$ and $D^0$ respectively.

All the model parameters $\Theta$ are learned by minimizing the combined loss. Following the representation in Figure \ref{fig:structure} (a) the total loss $L$ is defined as:
\begin{equation}
	L= Loss1+ Loss2 + Loss3,
	\label{eq:total_loss}
\end{equation}

Specifically, each term in Equation (\ref{eq:total_loss}) is denoted as
\begin{equation}
	\begin{cases}
		Loss1 &= \lambda_1 L_y, \\
		Loss2 &= \lambda_2 L_d^{\dagger}, \\
		Loss3 &= \alpha \lambda_3(L_d^{0}+(1-\alpha)L_d^{1}).
	\end{cases}
\label{eq:loss_cases}
\end{equation}

Based on Equation (\ref{eq:loss_cases}), the Equation (\ref{eq:total_loss}) can be denoted as
\begin{equation}
	L= \lambda_1 L_y + \lambda_2 L_d^{\dagger} + \alpha \lambda_3(L_d^{0}+(1-\alpha)L_d^{1}), \label{eq:loss_fn}
\end{equation}
where $\alpha$ is a hyperparameter that balances the importance of two intra-class domain confusion losses. In the experiment, the default value one is used for $\lambda_1$, $\lambda_2$ and $\lambda_3$ for our model. In the CTR prediction task, the importance of confusion loss in the two classes should be the same, and bias to either class will increase the prediction loss of the model. $\alpha$ is set to 0.5 and the validation set is verified on.

Generally, when $\lambda_1$, $\lambda_2$ and $\lambda_3$ are set to 1, it can be formalized that
\begin{equation}
	\begin{aligned}
		L(\theta_f,\theta_y,\theta_d^{\dagger},\theta_d^{1},\theta_d^0)=&\frac{1}{n}\sum_{i=1}^{n} L_y^i(\theta_f,\theta_y)+\frac{1}{n}\sum_{i=1}^{n} L_d^{\dagger i}(\theta_f,\theta_d^{\dagger})
		\\
		&+\alpha(\frac{1}{\tau}\sum_{i=1}^{\tau} L_d^{0i}(\theta_f,\theta_d^0))
		\\
		&+(1-\alpha)(\frac{1}{\rho}\sum_{i=\tau+1}^{n}L_d^{1i}(\theta_f,\theta_d^1)),
	\end{aligned}
\end{equation}
where $\theta_f, \theta_y, \theta_d^{\dagger}, \theta_d^{1}, \theta_d^0$ are parameters of feature extractor of the model (Domain Agnostic Layer and its former components), label predictor is demonstrated as $G$, global domain classifier is demonstrated as $D^{\dagger}$, and two intra-class domain classifiers is demonstrated as $D^{1}, D^0$ respectively. 

The optimization of the model can be realized by finding the saddle point $\theta_f,\theta_y,\theta_d^{\dagger},\theta_d^{1},\theta_d^0$ such that
\begin{equation}
	(\hat{\theta}_f,\hat{\theta}_y,\hat{\theta}_d^{\dagger},\hat{\theta}_d^{1},\hat{\theta}_d^0)=  \underset{\theta_f,\theta_y,\theta_d^{\dagger},\theta_d^{1},\theta_d^0}{{\arg\min} \, L}.
	\label{eq:loss_opt}
\end{equation}

Theoretically, the optimal solution defined by Equation (\ref{eq:loss_opt}) can be found as a stationary point of the following gradient updates:
\begin{equation}
	\begin{matrix}
		\theta_f^{(t+1)} \longleftarrow \theta_f^{(t)}-\mu(\frac{\partial L_y^i}{\partial \theta_f}+\frac{\partial L_d^{\dagger i}}{\partial \theta_f}+\alpha\frac{\partial L_d^{0i}}{\partial \theta_f} +(1-\alpha)\frac{\partial L_d^{1i}}{\partial \theta_f}), \\
		\theta_y^{(t+1)} \longleftarrow \theta_y^{(t)}-\mu\frac{\partial L_y^i}{\partial \theta_y} , \\
		\theta_d^{\dagger (t+1)} \longleftarrow \theta_d^{\dagger (t)}-\mu\frac{\partial L_d^{\dagger i}}{\partial \theta_d^{\dagger}}, \\
		\theta_d^{0 (t+1)} \longleftarrow \theta_d^{0 (t)}-\mu\times \alpha\frac{\partial L_d^{0 i}}{\partial \theta_d^0}, \\
		\theta_d^{1 (t+1)} \longleftarrow \theta_d^{1 (t)}-\mu\times (1-\alpha)\frac{\partial L_d^{1 i}}{\partial \theta_d^1},
	\end{matrix}
	\label{eq:grad1}
\end{equation}
where $\mu$ is the learning rate. Normally, the stochastic gradient descent (SGD) algorithm or Adam algorithm \citep{kingma2014adam} is used to optimize the loss function and then realize the learning of model parameters. However, it can be seen that from Equation (\ref{eq:grad1}), due to the difference in the model structure and scales of features, the gradient values of the parameters of different components of the model may be quite different in the optimization process, which is reflected in the trends of loss curves (\textit{Loss1} to \textit{Loss3}) in the joint training of multiple tasks, it will be covered in Section \ref{Experiments}. 

\begin{algorithm}[htb]
	\caption{Algorithm of DADIN}
	\label{alg:DADIN}
	\begin{algorithmic}[1]
		\Require source domain data, target domain data and hyper-parameters
		
		\Ensure DADIN $F$'s parameters
		
		\State Initialize $F$;
		\State \textbf{for} each epoch during the training \textbf{do}
		\State \quad Sample instances has feature from target domain or source domain;
		\State \quad Compute the embedding vector of user field: $p_k$ and the embedding vector of item filed: $q^s_j$ or $q^t_j$;
		\State \quad Employ \textit{Sequence Aggregation} to compute the compact feature representation of target history field: $a_j^k$ by Equation (\ref{eq:ag}) and the compact feature representation of source history field: $b_j^k$;
		\State \quad Employ \textit{Interest Attention} to to weighted concatenate four field vector ($p_k, q_j^s, a_j^k, b_j^k$ or $p_k, q_j^t, a_j^k, b_j^k$) as Equation (\ref{eq:interest_attention}) and (\ref{eq:cal_weight}): $m_{jk}$;
		\State \quad Employ \textit{Feature Crossing Layer} $f_{BI}$ and DNN to obtain the combination of features from different fields: $h_{spec}$;
		\State \quad Employ \textit{Domain Agnostic Layer} to compute domain-agnostic information: $h_{DA}$ and employ GRL on $h_{DA}$;
		\State \quad Employ \textit{Global Domain Classifier} $D^{\dagger}$ and two \textit{intra-class Domain Classifiers} $D^1, D^0$ on $h_{DA}$ to compute the domain predictions: $\hat d, \hat d^1, \hat d^2$ by Equations (\ref{eq:domain})-(\ref{eq:domain0});
		\State \quad Compute the compose of domain-agnostic information $h_{DA}$ and domain-related information $h_{spec}$: $h = h_{DA} + h_{spec}$;
		\State \quad Employ the \textit{Label Classifier} $G$ on $h$ to compute the label prediction: $\hat y$ by Equation (\ref{eq:predict});
		\State \quad Compute the total loss $L$ by Equation (\ref{eq:loss_fn}) and update $F$.
		\State \textbf{end} \textbf{for}
	\end{algorithmic}
\end{algorithm}

\section{Experiments}\label{Experiments}

The feasibility of DADIN is precisely verified through experiments. In Section \ref{ToyExp},three sets of conceptual experiments are first conducted on artificial data sets and the good performance of the model is intuitively verified on DA by visualizing the intermediate state of inputs under different model variants, which enhances the interpretability of the model. In Section \ref{ExpSetup}, the hyper-parameters and evaluation metrics of experiment results used are introduced in comparison experiments on real data sets. In Section \ref{Compare}, the basic information of the two real datasets we used are covered in detail, including data sources and basic statistical analysis of features, as well as how to preprocess the features of the dataset to make them conform to the requirements of our model input and how to resample the raw dataset to create a cold start situation. The method’s superior performance over the state-of-the-art single-domain and cross-domain recommendation algorithms is shown in a wide range of baselines \ref{Compare}. Besides, the validity of each component of the model is demonstrated in Section \ref{Ablation} by conducting an ablation study on various variants of the model. Finally, the trend of each loss function in the method (Section \ref{Analysis}) is analyzed to help understand the joint training process of the proposed models..

\subsection{Toy Experiments.}\label{ToyExp}
Toy experiment in imitation of \citep{hu2018conet} design is conducted for the purpose of verifying the effect of domain confusion loss realized by adding DA layer and domain discriminator. Similarly, a variant of \textit{inter-twinning moons} 2D problem is investigated. Among them, the distribution of target domain data is obtained by rotating the source domain data distribution 35 degrees around the center. According to the distribution, 300 source domain data and 300 target domain data are generated. To this end, DADIN-toy model for this problem is built separately, whose architecture only adds the designed domain adversarial module to the two-tower three-layers DNN. The results of the DADIN-toy on generated two-dimensional data compared with the naive DNN with the same two-tower structure and the same number of parameters verifies : \textbf{1)} Domain adversarial method can provide more accurate classification boundary for target domain. \textbf{2)} Domain confusion loss can confuse domain information to get invalid classification boundary of the domain classifiers. \textbf{3)} Domain agnostic layer can extract domain-agnostic features of data from both domains to enhance knowledge transfer from source domain to target domain.

\begin{figure}[ht]
	\centering
	\includegraphics[width=0.7\columnwidth]{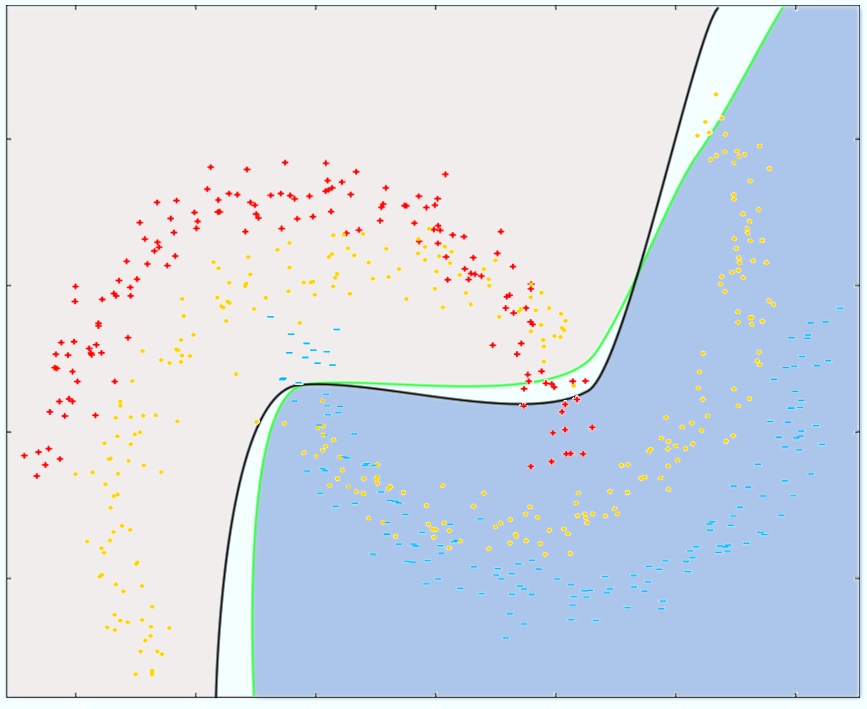}
	\caption{{\bf Verification results of DADIN and DNN on artificial datasets, in which red point "+" is the positive sample of the source domain, light blue point "-" is the negative sample of the source domain, yellow point is the target domain sample, light red part is the area judged as positive example, blue part is the area judged as negative example, black line is the classification boundary of DADIN, and green line is the classification boundary of DNN.}}\label{fig:toy_label}
\end{figure}

First, a test is conducted to figure out whether domain adversarial method can enhance target domain discrimination. In this stage, all 300 source domain data and half 150 target domain data are used to train classifiers DADIN and DNN, respectively, to simulate a real cold start. The obtained model is verified on all the target domain data, and its visual results are shown in Figure \ref{fig:toy_label}.

Figure \ref{fig:toy_label} shows that the classification boundary of DADIN is more accurate and more consistent with the distribution of data in the target domain. On the contrary, although the classification boundary of DNN seems to fit well, it can be seen in details that it does not completely understand the data distribution of the target domain, but only overfits the training data. This problem is caused by the failure to use the source domain data for the training of the target domain classifier. The cross-domain information is not fully utilized and the only marginal distribution is aligned in the embedding space, but the conditional distribution of the target domain and the source domain is not learned. Therefore, the proposed domain adversarial method can enhance the target domain classification by conditional distribution alignment.

\begin{figure}[ht]
	\centering
	\subfloat[{\bf The domain discrimination classification boundary of DADIN, almost all samples are classified as negative cases.}]{\includegraphics[width=0.47\columnwidth]{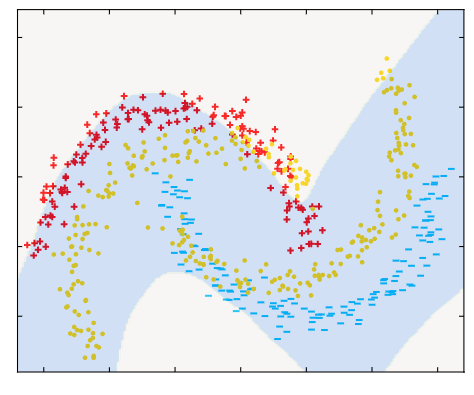}}\hspace{5pt}
	\hfill
	\subfloat[{\bf The domain discrimination classification boundary of DNN, and about half of the samples can be correctly classified.}]{\includegraphics[width=0.47\columnwidth]{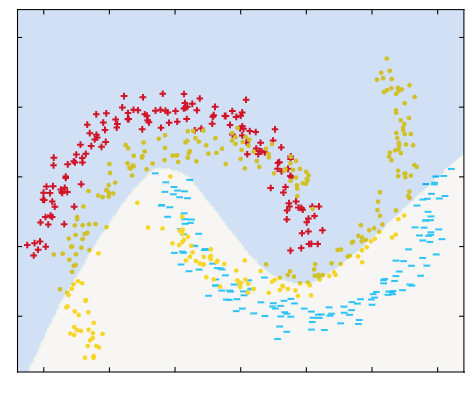}}
	\caption{{\bf The domain discrimination classification boundaries of DADIN and DNN.}}
	\label{fig:toy_domain}
\end{figure}

Secondly, it is discussed whether the domain confusion loss corresponding to the proposed method can confuse the domain information to get the invalid boundary for domain discrimination. Accordingly, all the data are used to train the DADIN and a DNN for domain discrimination task, and the domain classifiers in DADIN and the DNN are compared. The visualization results is shown in Figure \ref{fig:toy_domain}.

Figure \ref{fig:toy_domain} (b) shows that DNN has learned the pattern of source and target domain (the target domain distribution is obtained by rotating the source domain distribution by 35 degrees), but the classification boundary is still not accurate enough. This is because there is no structure to completely separate domain-related information and domain-agnostic information in DNN. When extracting the information of source and target domain, DADIN adopts the method of separation of domain-related and domain-agnostic information. It adds skip-connection-based domain agnostic layer and class confusion loss to specialize domain information. Figure \ref{fig:toy_domain} (b) demonstrates that the domain-agnostic information $h_{DA}$ totally confuse the domain classifiers of DADIN.

In addition, in order to further study the role that domain agnostic layer in the model plays in extracting domain-agnostic features of data from both domains, DADIN-toy is trained, and all sample data from both domains are used. The technology of dimensional reduction visualization is used to show the distribution of the intermediate state $\mathbf{h}_{spec}$ of the input data before the domain agnostic layer transformation and the intermediate state $\textbf{h}_{DA}$ after the domain agnostic layer transformation. Specifically, the principal component analysis method (PCA) is used to transform the intermediate state $\mathbf{h}_{DA},\mathbf{h}_{spec} \in \mathbb{R}^{d}$ of the forward process of the model into a two-dimensional vector, and represent the positive and negative case samples from different fields with different color points in the plane coordinate system. In this way, their distribution can be observed more intuitively.

\begin{figure}[ht]
	\centering
	\subfloat[{\bf The intermediate state $\textbf{h}_{spec}$ of the input data before the domain agnostic layer transformation (after PCA reduction). The red point "+" is a positive example from the source domain, the dark blue dot is a positive example from the target domain.}]{\includegraphics[width=0.47\columnwidth]{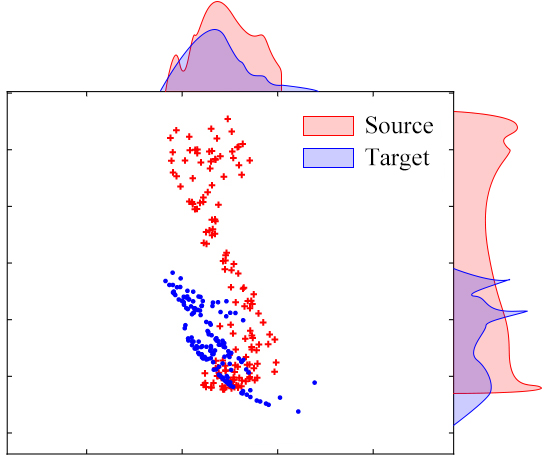}} \hspace{5pt}
	\hfill
	\subfloat[{\bf The intermediate state $\textbf{h}_{DA}$ of the input data after the domain agnostic layer transformation (after PCA reduction). The red point "+" is a positive example from the source domain, the dark blue dot is a positive example from the target domain.}]{\includegraphics[width=0.47\columnwidth]{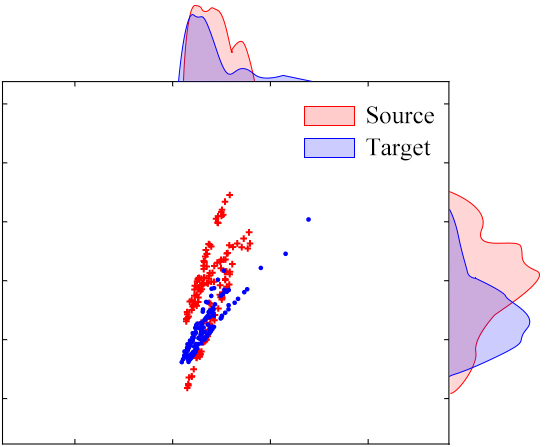}}
	\caption{{\bf The visualization of the intermediate state $\mathbf{h}_{spec}$ and $\textbf{h}_{DA}$'s distribution. The blue and red probability density function curves around the Figure \ref{fig:toy_pca} (a) and (b) represent the distribution of target domain samples and source domain samples after dimensionality reduction respectively.}}
	\label{fig:toy_pca}
\end{figure}

Figure \ref{fig:toy_pca} shows that compared with $\mathbf{h}_{spec}$, the positive samples of the source and target domains of  $\mathbf{h}_{DA}$ after domain agnostic layer transformation (red point "+" and blue dots) are more closely aligned in two-dimensional space, which indicates that the domain agnostic layer of the model can extract domain-agnostic features from samples in the source and target domain, so as to achieve the marginal distribution and conditional distribution alignment of the target and source domain samples in the feature space. To some extent, it verifies the validity of our domain agnostic layer.

So far, the feasibility of the proposed domain adversarial approach has been demonstrated using artificial datasets for both three purposes.

\subsection{Experiments Setup}\label{ExpSetup}

Before introducing the real dataset experiments, it is necessary to explain some details of the experiment from below aspects : \textbf{1)} dataset selection, \textbf{2)} cold start scenario construction (by resampling method), \textbf{3)} independent repeated experiment settings.

\textbf{Dataset}\quad The model is evaluated and compared with other models on two large real-world datasets. The first is the dataset of Huawei's 2022 cross-domain CTR prediction competition \footnote{https://www.huawei.com/}, which is large in scale and has relatively complete features. The user click news data is from the source domain, and the user click advertisement data is from the target domain. In addition, the user set of target domain in the original dataset is included in the user set of source domain, which is convenient to build a cold start scenario. Secondly, two fields ’music’ and ’movie’ are selected from Amazon dataset \footnote{http://jmcauley.ucsd.edu/data/amazon/index\_2014.html} and the overlapped part of users is captured as the total dataset. The user behavior data of 'movie' with a large amount of data is taken as the source domain and the user behavior data of 'music' is taken as the target domain. In addition, historical behavior statistics are carried out for the data of the two domains to generate historical behavior features.

\textbf{Resampling}\quad In order to simulate the real cold start environment, a special resampling algorithm is designed to resample the data. The main idea is to make the user set of the training set in the target domain and the user set of the verification and test set not overlap as much as possible after resampling, and the source domain contains all the user sets of the target domain. The whole dataset is divided according to the user set, and the training set, verification set and test set are obtained respectively.

\begin{algorithm}[htb]
	\caption{Algorithm of resampling}
	\label{alg:resampling}
	\begin{algorithmic}[1]
		\Require $A_{train}$, $A_{cross}$
		
		\Ensure \textbf{train\_data}, \textbf{valid\_data}, \textbf{test\_data}
		
		\State Take the ads field user (Huawei has too much data and has resamped an 50\% of all data additionally) as all the users to be sampled and write them as \textbf{user\_list};
		\State $A_{train}$ 
		of the users are taken as all the users in the training set and recorded as \textbf{user\_train\_list}. The users contained in \textbf{user\_train\_list} are used to obtain data on the ads field and feeds field respectively, and are recorded as \textbf{train\_data};
		\State  Sample $A_{cross}$ of the user in \textbf{user\_train\_list} as users that overlap the validation set and test set. And take out the last day records of the overlap users from the ads data in the \textbf{train\_data} as the validation set (50\%), namely \textbf{valid\_data} and test set (50\%) namely \textbf{test\_data}, and delete this part of data in the training set;
		\State Divide the remaining $1-A_{train}$ of the users in \textbf{user\_list} into test set and validation set, and take out the last day data of ads data according to these users and add them to the \textbf{test\_data} and \textbf{validation\_data} obtained in step 3: respectively;
		\State Finish the sampling. Return \textbf{train\_data}, \textbf{valid\_data}, \textbf{test\_data}.
	\end{algorithmic}
\end{algorithm}

\textbf{Evaluation Metrics}\quad \textbf{1)} AUC: $$AUC = \frac{\sum pred_{pos} > pred_{neg}}{posistiveNum \times negativeNum},$$ where the denominator $posistiveNum \times negativeNum$ is the total number of combinations of positive and negative samples and the numerator $\sum pred_{pos} > pred_{neg}$ is the number of combinations where positive samples are greater than negative samples;
\textbf{2)} Logloss: $$LogLoss=-\frac{1}{n}\sum_i[y_i \log(\hat y_i) + (1-y_i)\log(1-\hat y_i)]$$ where $n$ is the number of instances, $y_i$ is the true label of the instance $i$, $\hat y_i$ is the CTR prediction of the instance $i$.

\textbf{Independent repeat experiment}\quad In order to make the experiment results more reliable, different random seeds for each dataset are used to generate 10 sets of data. These three sets of data are used to compare the models (random seeds of the models and random seeds related to experiment itself remain unchanged), and finally record the mean and standard deviation of the evaluation results, so that the results are more reliable.

\subsection{Comparing Different Approaches}\label{Compare}

Both single-domain and cross-domain methods are compared. Specifically, when organizing the input of the single-domain model, the instance of the two domains is directly concatenated with the attribute "user\_Id" as the key, and the missing feature is filled in with the special token "na\_value" as a category in the categorical variable. The concatenated instance is input into the single-domain model to give the corresponding CTR prediction, and the loss is calculated with the true labels in its domain. For the cross-domain methods, the input is organized for different cross-domain models in the same way as in their original paper; For some two-stage models such as MV-DNN, the feature extraction part and modified the second-stage is retained by changing loss function of the model to adapt it to the cross-domain CTR prediction task. When reproducing these baselines, the model structure and hyperparameter settings used in their original papers are followed as much as possible to ensure the authenticity and validity of the comparison experiments.

\subsubsection{Baseline}

\begin{table}[]
	\centering
	\caption{{\bf Baseline models}}
	\resizebox{\textwidth}{!}{%
		\begin{tabular}{|c|l|l|l|}
			\hline
			\multicolumn{1}{|l|}{}                  & \multicolumn{1}{c|}{\textbf{abbr}} & \multicolumn{1}{c|}{\textbf{full name}}  & \multicolumn{1}{c|}{\textbf{description}}                                                                                                                                                                                                          \\ \hline
			\multirow{13}{*}{\textbf{Single-domain}} & LR                                 & Logistic Regression                      & \begin{tabular}[c]{@{}l@{}}A linear model with Sigmoid function as the activation function.\end{tabular}                                                                                                               \\ \cline{2-4} 
			& FM                                 & Factorization Machine                    & \begin{tabular}[c]{@{}l@{}}Advanced LR model that learns the weights of second-order feature \\ interactions in a way of latent vector learning.\end{tabular}                                                                                                          \\ \cline{2-4} 
			& Wide\&Deep                         & Wide\&Deep                               & \begin{tabular}[c]{@{}l@{}}Classic deep learning framework for CTR prediction consisting of\\ a linear model as wide part and a DNN as deep part.\end{tabular}                                                                                  \\ \cline{2-4} 
			& xDeepFM                            & eXtreme Deep Factorization Machine       & \begin{tabular}[c]{@{}l@{}}Powerful CTR prediction model based on Wide\&Deep framework,\\ using Compressed Interaction Network (CIN) to explicitly model \\ higher-order crossing features in a vector-wise manner.\end{tabular}                                                 \\ \cline{2-4} 
			& DIN                                & Deep Interest Network                    & \begin{tabular}[c]{@{}l@{}}DIN assigns activation weights to the historical behavior vectors\\according to the candidate items and combine them as users’ interest\\ given the candidate item by an attention-based pooling method.\end{tabular} \\ \cline{2-4} 
			& NFM                                & Neural Factorization Machine             & \begin{tabular}[c]{@{}l@{}}NFM introduces the Bi-Interaction Pooling layer to replace the \\concatenation layer of classic DNN to add enough feature interaction \\ information to the bottom layer.\end{tabular}                                   \\ \hline
			\multirow{10}{*}{\textbf{Cross-domain}}  & MV-DNN                             & Multi-View DNN                           & \begin{tabular}[c]{@{}l@{}}MV-DNN maps the information from the user side and the item side \\ into a same embedding space through a shared embedding layer, then\\ models their interactions.\end{tabular}                                              \\ \cline{2-4} 
			& MLP++                              & MLP++                                    & \begin{tabular}[c]{@{}l@{}}The naive version of CoNet modeling representation of users and \\items separately with two independent MLP.\end{tabular}                                                                                              \\ \cline{2-4} 
			& CSN                                & Cross-Stitch Network                     & \begin{tabular}[c]{@{}l@{}}CSN extracts the features of images in the two fields independently \\ and achieves the goal of bidirectional knowledge transfer by linear \\ combination of feature maps.\end{tabular}                                    \\ \cline{2-4} 
			& CoNet                              & Collaborative cross Network              & \begin{tabular}[c]{@{}l@{}}CoNet introduces cross connection units that changes linear\\ combination to linear transformation realizing more fine-grained \\ and sparse knowledge  transfer.\end{tabular}                                             \\ \cline{2-4} 
			& MiNet                              & Mixed Interest Network                   & \begin{tabular}[c]{@{}l@{}}MiNet jointly models user interest from multiple domains, weighting\\ them in a hierarchical attention-based manner.\end{tabular}                                                                                          \\ \cline{2-4} 
			& DADIN                              & Domain Adversarial Deep Interest Network & Domain Adversarial Deep Interest Network proposed in this paper.                                                                                                                                                                                   \\ \hline
		\end{tabular}%
	}
\end{table}

\textbf{Baseline hyperparameter settings}\quad Considering the equipment limitations and model performance, the different hyperparameters selected for the model in the comparison experiment are shown in the table \ref{tab:parameters}. Adam \citep{richardson2007predicting} is used as the optimizer for all models, the embedding dimension of all models is fixed as 64, and early stopping is used in training to obtain the optimal model. Moreover, cross entropy loss is used as the loss function for both single-domain models and cross-domain models.
\begin{table}[t]
	\centering
	\caption{{\bf Baseline hyperparameter settings.}}
	\label{tab:parameters}
	\resizebox{0.65\textwidth}{!}{%
		\begin{tabular}{|cc|c|c|c|}
			\hline
			\multicolumn{2}{|c|}{\textbf{Model}}                                       & \textbf{batch size} & \textbf{hidden units}     & \textbf{learning rate} \\ \hline
			\multicolumn{1}{|c|}{\multirow{6}{*}{\textbf{Single-domain}}} & LR         & \multirow{2}{*}{1000}                & \multirow{2}{*}{-}                         & \multirow{2}{*}{1.00E-03}               \\
			\multicolumn{1}{|c|}{}                               & FM         &                                      &                                            &                                         \\ \cline{2-5}
			\multicolumn{1}{|c|}{}                               & Wide\&Deep & \multirow{4}{*}{1000}                & \multirow{4}{*}{[512,512,512]} & \multirow{4}{*}{1.00E-03}               \\
			\multicolumn{1}{|c|}{}                               & xDeepFM    &                                      &                                            &                                         \\
			\multicolumn{1}{|c|}{}                               & DIN        &                                      &                                            &                                         \\
			\multicolumn{1}{|c|}{}                               & NFM        &                                      &                                            &                                         \\ \hline
			\multicolumn{1}{|c|}{\multirow{7}{*}{\textbf{Cross-domain}}}  & MV-DNN     & \multirow{7}{*}{2000}                & \multirow{7}{*}{[512,128,64]}  & \multirow{7}{*}{1.00E-04}               \\
			\multicolumn{1}{|c|}{}                               & MLP++      &                                      &                                            &                                         \\
			\multicolumn{1}{|c|}{}                               & CSN        &                                      &                                            &                                         \\
			\multicolumn{1}{|c|}{}                               & CoNet      &                                      &                                            &                                         \\
			\multicolumn{1}{|c|}{}                               & MiNet      &                                      &                                            &                                         \\
			\multicolumn{1}{|c|}{}                               & DADIN8     &                                      &                                            &                                         \\
			\multicolumn{1}{|c|}{}                               & DADIN++    &                                      &                                            &                                         \\ \cline{1-5}
		\end{tabular}%
	}
\end{table}

\subsubsection{Results}

In this section, the recommendation performance of different methods is discussed. Table \ref{tab:comparsion} shows the results of different models on the two datasets under two evaluation metrics. The relative improvement of the model over other baselines is given under the row of AUC. Specifically, 10 different random seeds are set up to conduct Independent repetition experiments 10 times on two datasets for the model and all the competitors, then the mean and standard deviation of two evaluation metrics are calculated for each model as the final result in Table \ref{tab:comparsion}. The number of parameters of each model is also provided. The model outperforms a bunch of baseline models on two datasets without increasing the computational complexity of the model.

\begin{table*}[]
	\centering
	\caption{{\bf Cross-domain CTR prediction performance on the two real datasets of the proposed method and baselines.}}
	\label{tab:comparsion}
	\resizebox{\tblwidth}{!}{%
		\begin{tabular}{|l|cc|c|c|c|c|}
			\hline
			\textbf{Dataset}                                        & \multicolumn{2}{c|}{\textbf{Model}}                                             & \textbf{AUC}                  & \textbf{improve of AUC(\%)} & \textbf{Logloss} & \textbf{\#Parameters}   \\ \hline
			\multicolumn{1}{|c|}{\multirow{12}{*}{\textbf{Amazon}}} & \multicolumn{1}{c|}{\multirow{6}{*}{Single-domaim}} & LR               & 0.50862($\pm$ 0.008) & 27.07\%            & 0.57614 & \textbf{0.22}M \\
			\multicolumn{1}{|c|}{}                         & \multicolumn{1}{c|}{}                               & FM               & 0.51930($\pm$ 0.002) & 25.54\%            & 1.19791 & 14.43M         \\
			\multicolumn{1}{|c|}{}                         & \multicolumn{1}{c|}{}                               & Wide\&Deep       & 0.49617($\pm$ 0.004) & 28.85\%            & 1.47038 & 15.18M         \\
			\multicolumn{1}{|c|}{}                         & \multicolumn{1}{c|}{}                               & xDeepFM          & 0.50537($\pm$ 0.008) & 27.53\%            & 1.51702 & 14.47M         \\
			\multicolumn{1}{|c|}{}                         & \multicolumn{1}{c|}{}                               & DIN              & 0.50138($\pm$ 0.004) & 28.11\%            & 2.51903 & 14.98M         \\
			\multicolumn{1}{|c|}{}                         & \multicolumn{1}{c|}{}                               & NFM              & 0.57002($\pm$ 0.005) & 18.26\%            & 1.94616 & 14.99M         \\ \cline{2-7} 
			\multicolumn{1}{|c|}{}                         & \multicolumn{1}{c|}{\multirow{6}{*}{Cross-domain}}  & MV-DNN           & 0.62652($\pm$ 0.035) & 10.16\%            & 0.62100 & 30.43M         \\
			\multicolumn{1}{|c|}{}                         & \multicolumn{1}{c|}{}                               & MLP++            & 0.66092($\pm$ 0.037) & 5.23\%             & 0.53870 & 4.58M          \\
			\multicolumn{1}{|c|}{}                         & \multicolumn{1}{c|}{}                               & CSN              & 0.66714($\pm$ 0.032) & 4.34\%             & 0.52994 & 4.58M          \\
			\multicolumn{1}{|c|}{}                         & \multicolumn{1}{c|}{}                               & CoNet            & 0.66674($\pm$ 0.029) & 4.39\%             & \textbf{0.52932} & 4.58M          \\
			\multicolumn{1}{|c|}{}                         & \multicolumn{1}{c|}{}                               & MiNet            & 0.69240($\pm$ 0.046) & 0.71\%             & 0.55484 & 4.49M          \\
			\multicolumn{1}{|c|}{}                         & \multicolumn{1}{c|}{}                               & \textbf{DADIN++} & \textbf{0.69738($\pm$ 0.008)} & -                  & 0.57720 & 4.61M          \\ \hline
			\multirow{12}{*}{\textbf{Huawei}}                       & \multicolumn{1}{c|}{\multirow{6}{*}{Single-domaim}} & LR               & 0.72240($\pm$ 0.008) & 7.29\%             & 0.10116 & \textbf{0.92}M \\
			& \multicolumn{1}{c|}{}                               & FM               & 0.69917($\pm$ 0.027) & 10.27\%            & 0.12173 & 59.46M         \\
			& \multicolumn{1}{c|}{}                               & Wide\&Deep       & 0.76328($\pm$ 0.015) & 2.04\%             & 0.08665 & 61.19M         \\
			& \multicolumn{1}{c|}{}                               & xDeepFM          & 0.74465($\pm$ 0.009) & 4.44\%             & 0.10059 & 59.59M         \\
			& \multicolumn{1}{c|}{}                               & DIN              & 0.76226($\pm$ 0.016) & 2.18\%             & 0.08670 & 60.23M         \\
			& \multicolumn{1}{c|}{}                               & NFM              & 0.73720($\pm$ 0.009) & 5.39\%             & 0.10241 & 59.95M         \\ \cline{2-7} 
			& \multicolumn{1}{c|}{\multirow{6}{*}{Cross-domain}}  & MV-DNN           & 0.71864($\pm$ 0.001) & 7.77\%             & 0.07829 & 40.24M         \\
			& \multicolumn{1}{c|}{}                               & MLP++            & 0.77510($\pm$ 0.003) & 0.53\%             & 0.07326 & 14.39M         \\
			& \multicolumn{1}{c|}{}                               & CSN              & 0.77859($\pm$ 0.006) & 0.08\%             & 0.07276 & 14.39M         \\
			& \multicolumn{1}{c|}{}                               & CoNet            & 0.77186($\pm$ 0.007) & 0.94\%             & 0.07345 & 14.39M         \\
			& \multicolumn{1}{c|}{}                               & MiNet            & 0.76363($\pm$ 0.004) & 2.00\%             & 0.07443 & 14.30M         \\
			& \multicolumn{1}{c|}{}                               & \textbf{DADIN++} & \textbf{0.77921($\pm$ 0.003)} & -                  & \textbf{0.07242} & 14.42M         \\ \hline
		\end{tabular}%
	}
\end{table*}

\textbf{Single-domain Methods}\quad For each single-domain model, the performance on datasets with different data amounts is worse than that of the cross-domain model (except MV-DNN). In the Huawei dataset with a large amount of data, the AUC of some models (such as Wide\&Deep) outnumbers that of MV-DNN and is close to MLP++ and MiNet. This is because these single-domain models are designed with better feature extraction modules (Wide\&Deep, xDeepFM, NFM) or exploit user's historical behaviour information (DIN), which can better learn the distribution of target domains on large datasets. However, in the Amazon dataset with a small amount of data, the effect of the single-domain model is inferior to that of the cross-domain model, which is an inevitable result caused by too little information of the distribution of the target domain.

\textbf{Cross-domain Methods}\quad Among a set of cross-domain CTR prediction models, MV-DNN with two-stage modeling has the worst result on the two real datasets, which indicates that the one-stage end-to-end CTR prediction model has the advantage in the cross-domain CTR prediction task. In Huawei and Amazon datasets, there is no significant difference in the performance of CoNet, CSN, and MLP++, which demonstrates that there is no significant difference in the shared cross connection coefficient of the intermediate state of the two-tower base network with matrix $\mathbf{H}_{D}$ or scalar $\alpha_D$. It is worth noting that the performance of MiNet in Huawei dataset is slightly lower than that of CoNet series models, while the situation is the reverse in Amazon dataset. Considering the difficulty of knowledge transfer between movie ratings data and music ratings data in Amazon dataset, it is believed that the more fine-grained operation of different field-aware embedding can enhance the transfer of knowledge between two far distributed fields, as MiNet has done. Compared with the most competitive baselines in both datasets, the DADIN model has a significant improvement, which is 0.08\% higher than CSN in Huawei dataset and 0.71\% higher than MiNet in Amazon dataset.

\subsection{Ablation Study}\label{Ablation}

In this section, in order to quantitatively verify the validity of each component of our proposed model, 10 variants of the DADIN model are examined and experiments are conducted on two real data sets. To ensure comparability of experimental results, all of the model variants have the same hyperparameter settings. The 10 variants are: \textbf{1)} DADIN1: Replace item-level interest attention-based sequence aggregation with average pooling, \textbf{2)} DADIN2: Replace interest-level attention-based concatenate with equal weight concatenation, \textbf{3)} DADIN3: Remove Bi-interaction Pooling layer, \textbf{4)} DADIN4: Remove DNN layers, \textbf{5)} DADIN5: Remove both Bi-interaction layer and DNN layers, or remove feature crossing part, \textbf{6)} DADIN6: Remove domain agnostic layer, \textbf{7)} DADIN7: Remove global domain confusion loss $L_d^{\dagger}$, \textbf{8)} DADIN8: Remove intra-class domain confusion loss $L_d^0$ and $L_d^{1}$, \textbf{9)} DADIN9: Remove both global domain confusion loss and intra-class domain confusion loss, that is the non-adversarial version of the model, \textbf{10)} DADIN++ : The complete version of the model.

\begin{table*}[t]
	\centering
	\caption{{\bf The comparative results of DADIN++ and all its variants.}}
	\resizebox{0.9\textwidth}{!}{%
		\begin{tabular}{|c|ccc|ccc|}
			\hline
			& \multicolumn{3}{c|}{\textbf{Amazon}}                                                                                                     & \multicolumn{3}{c|}{\textbf{Huawei}}                                                                                                     \\ \hline
			\textbf{Model}   & \multicolumn{1}{c|}{AUC}               & \multicolumn{1}{c|}{improve of AUC(\%)} & \multicolumn{1}{c|}{LogLoss}           & \multicolumn{1}{c|}{AUC}               & \multicolumn{1}{c|}{improve of AUC(\%)} & \multicolumn{1}{c|}{LogLoss}            \\ \hline
			DADIN\_1         & \multicolumn{1}{c|}{0.67336}          & \multicolumn{1}{c|}{3.23\%}             & \multicolumn{1}{c|}{0.63213}           & \multicolumn{1}{c|}{0.78174}          & \multicolumn{1}{c|}{0.83\%}             & \multicolumn{1}{c|}{0.07012}           \\
			DADIN\_2         & \multicolumn{1}{c|}{0.66605}           & \multicolumn{1}{c|}{4.28\%}             & \multicolumn{1}{c|}{0.69786}           & \multicolumn{1}{c|}{0.78012}          & \multicolumn{1}{c|}{1.03\%}             & \multicolumn{1}{c|}{0.07078}           \\
			DADIN\_3         & \multicolumn{1}{c|}{0.53549}          & \multicolumn{1}{c|}{23.04\%}            & \multicolumn{1}{c|}{\textbf{0.60213}}  & \multicolumn{1}{c|}{0.78181}          & \multicolumn{1}{c|}{0.82\%}             & \multicolumn{1}{c|}{0.06884}           \\
			DADIN\_4         & \multicolumn{1}{c|}{0.69117}          & \multicolumn{1}{c|}{0.67\%}             & \multicolumn{1}{c|}{0.72632}           & \multicolumn{1}{c|}{0.75590}          & \multicolumn{1}{c|}{4.11\%}             & \multicolumn{1}{c|}{0.07154}            \\
			DADIN\_5         & \multicolumn{1}{c|}{0.64947}          & \multicolumn{1}{c|}{6.66\%}             & \multicolumn{1}{c|}{0.84493}           & \multicolumn{1}{c|}{0.75744}           & \multicolumn{1}{c|}{3.91\%}             & \multicolumn{1}{c|}{0.07187}           \\
			DADIN\_6         & \multicolumn{1}{c|}{0.51749}          & \multicolumn{1}{c|}{25.63\%}            & \multicolumn{1}{c|}{2.80954}           & \multicolumn{1}{c|}{0.50000}               & \multicolumn{1}{c|}{36.57\%}            & \multicolumn{1}{c|}{0.24745}           \\
			DADIN\_7         & \multicolumn{1}{c|}{0.65085}           & \multicolumn{1}{c|}{6.46\%}             & \multicolumn{1}{c|}{0.62219}            & \multicolumn{1}{c|}{0.77122}          & \multicolumn{1}{c|}{2.16\%}             & \multicolumn{1}{c|}{0.07112}           \\
			DADIN\_8         & \multicolumn{1}{c|}{0.59945}          & \multicolumn{1}{c|}{13.85\%}            & \multicolumn{1}{c|}{0.65974}           & \multicolumn{1}{c|}{0.78230}          & \multicolumn{1}{c|}{0.76\%}             & \multicolumn{1}{c|}{0.07015}           \\
			DADIN\_9         & \multicolumn{1}{c|}{0.57983}          & \multicolumn{1}{c|}{16.67\%}            & \multicolumn{1}{c|}{0.90197}           & \multicolumn{1}{c|}{0.76980}            & \multicolumn{1}{c|}{2.34\%}             & \multicolumn{1}{c|}{0.07215}           \\
			\textbf{DADIN++} & \multicolumn{1}{c|}{\textbf{0.69581}} & \multicolumn{1}{c|}{-}                  & \multicolumn{1}{c|}{0.61499}           & \multicolumn{1}{c|}{\textbf{0.78828}} & \multicolumn{1}{c|}{-}                  & \multicolumn{1}{c|}{\textbf{0.06799}}  \\ \hline
		\end{tabular}%
		\label{tab:ablation_study}
	}
\end{table*}

In Table \ref{tab:ablation_study}, the comparative results of our model and all its variants are presented. The evaluation metrics include the AUC and Logloss mentioned in Section \ref{ExpSetup}. Obviously, the complete version of DADIN has higher AUC values on both datasets than its variants. The relative value of improvement of DADIN++ are calculated in the AUC indicator to other variants in Table \ref{tab:ablation_study}. 


In the Huawei dataset, the most significant improvement occurs in DADIN++ versus DADIN6 (36.57\%), where the domain agnostic layer is removed and DADIN++ reaches an AUC of 0.788, while DADIN6 has an AUC of 0.5. This means that after the domain agnostic layer is removed, the model collapses during the learning process, which confirms the importance of the domain agnostic layer. The understanding of this result is as follows: In DADIN6 with the domain agnostic layer removed, the constraint of domain confusion loss ensures that the intermediate state of the label predictor, global domain classifier and intra-class domain classifier's input is actually the domain-agnostic information $\textbf{h}_{DA}$. However, the experiments of DADIN6 has shown that only $\textbf{h}_{DA}$ is used to predict the CTR on the target domain results in model collapse, which means that the output of the label predictor is meaningless. In this process, from the perspective of multi-task learning, domain confusion loss optimization has a negative effect on the optimization of CTR prediction loss. From the perspective of domain adaptation, DADIN6 only realizes the marginal distribution alignment of the two domain's data in the feature space, ignoring the conditional distribution alignment. By introducing skip-connection-based domain agnostic layer, both marginal and conditional distribution alignments are achieved, enhancing the robustness of the model. The improvement of DADIN++ is 3.91\% compared with that of DADIN5. Since there are a large number of features from different fields in the Huawei dataset, it is necessary to use stacked feature crossing modules to carry out explicit and implicit feature crossing in a serial manner. Compared with the non-adversarial version of DADIN9, DADIN++ has been promoted to 2.34\% on the AUC, because the ads click instances from the target field and the news click instances from the source field in the Huawei dataset are close in their distribution. Therefore, the scheme using domain adaption based on domain adversarial learning to enhance cross-domain knowledge transfer has no obvious gain on this dataset.

In the Amazon dataset, due to the bias of converting the item score data into a click-or-not binary label, and the natural difficulty of transferring knowledge from the movie rating records to the music rating records, the DADIN model and its variants have varying degrees of decline in the absolute value of AUC compared with the Huawei dataset. However, the DADIN++ model still achieves an AUC close to 0.7, which is an acceptable result. Consistent with the results in Huawei dataset, compared with DADIN6 without domain agnostic layer and DADIN3 without explicit feature crossing module, the AUC of DADIN++ is greatly improved 25.63\% and 23.04\% respectively. It is worth noting that due to the far distance in the distribution of the film rating records in the source domain and the music rating records in the target domain, the DADIN++ has a big improvement compared with the non-adversarial version of DADIN9 (16.67\%). This verifies that our scheme of domain adaptation through domain adversarial learning to enhance cross-domain knowledge transfer very important when the distance between the target domain and the source domain is far away, which makes sure the effectiveness and scalability of the method.

The enhancement of attention-based sequence aggregation and attention-based weighted concatenation to the model is illustrated in the results of DADIN++ v.s. DADIN1 and DADIN++ v.s. DADIN2, which is slightly more significant on the Amazon dataset than Huawei dataset. On Amazon dataset,  the distance of distribution of data in the source domain is far from that in the target domain. In this case, the model should be able to more comprehensively capture user interest in the current project. For example, attention-based sequence aggregation is used to obtain a better feature expression of user's historical behavior (DADIN++ v.s. DADIN1). Or assign different importance to the features of the user field according to the correlation of the embedding vector when the given item  $i_j$ is given, so as to realize attention-based weighted concatenation (DADIN++ v.s. DADIN2). The concatenation embedding vector thus obtained can better adapt to the task of CTR prediction task. In order to verify the weights in the attention-based weighted concatenation (Equation (\ref{eq:interest_attention})) adopted in DADIN++ are meaningful instead of being equal in Equation (\ref{eq:without_interest_attention}), the weights $v_u,v_t,v_s$ are recorded and visualized in the forward process of the model when testing on the test set.

\begin{figure}[ht]
	\centering
	\includegraphics[width=0.45\columnwidth]{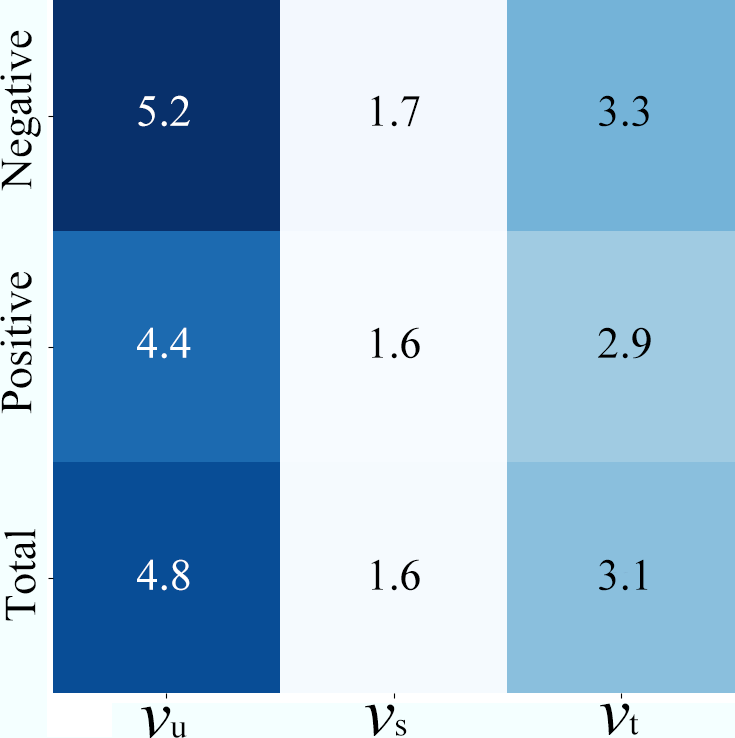}
	\caption{{\bf The mean of interest-level attention weights of positive and negative samples, and the mean of the means. The darker the color, the larger the value. The figure shows the specific values of three kinds of weights $v_u,v_s$ and $v_t$.}}\label{fig:attention_weight}
\end{figure}

Figure \ref{fig:attention_weight} shows that the interest-level attention weights of positive and negative samples in the test set are all greater than 1, and the embedding vectors from different fields have different weights, which verifies the effectiveness of our method. That is different from the equal weight embedding concatenation. In the comparison of the three embedding weights $v_u,v_s,v_t$, no matter the negative or positive samples, the result is $v_u>v_s>v_t$. This demonstrates that in the model, the feature $\mathbf{v}_i^{up} $from user profile is more important to CTR prediction task than the feature $\mathbf{v}_i^{th},$ rom user target history field. The feature $\mathbf{v}_i^{sh}$ of the user source history field is the least important. The mean values of the three weights $v_u,v_s,v_t$ of the negative samples are all greater than the mean values of the positive samples, indicating that for the positive samples, the contribution of the information in the user part to the final CTR prediction is less than that of the information in the item part. That is to say, when the click behavior occurs, the candidate items themselves tend to have some attractive characteristics. If a user does not click on the item, it is determined by the user's own characteristics such as occupation, gender, and preferences reflected in historical behavior.

\subsection{Analysis of loss functions}\label{Analysis}

\begin{figure*}[ht]
	\centering
	\includegraphics[width=\textwidth]{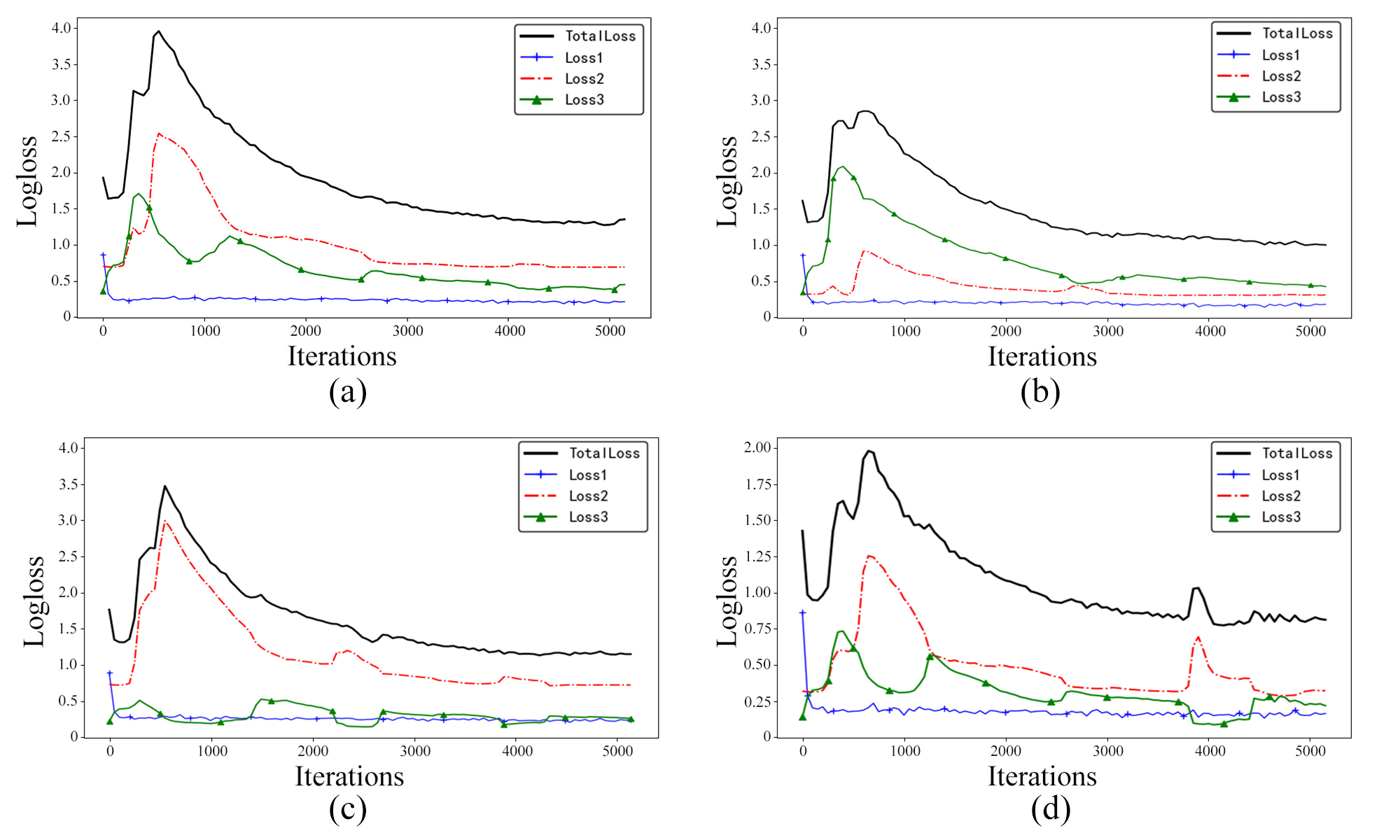}
	\caption{{\bf (a)-(d) show the variation of each part of the loss during the training process when different $\lambda_2, \lambda_3$ are selected. (a) $\lambda_2=1.0$, $\lambda_3=1.0$, $\text{AUC:}0.78343$. (b) $\lambda_2=0.5$, $\lambda_3=1.0$, $\text{AUC:}0.77695$. (c) $\lambda_2=1.0$, $\lambda_3=0.5$, $\text{AUC:}0.77712$. (d) $\lambda_2=0.5$, $\lambda_3=0.5$, $\text{AUC:}0.77665$.}}\label{fig:loss_study}
\end{figure*}

In order to better understand the multi-task joint training of the model and verify the convergence of the model, the total loss $L$, CTR prediction loss $L_ y$(\textit{Loss1}), global domain confusion loss (\textit{Loss2}) and intra-class domain confusion loss $\alpha L_d^{0}+(1-\alpha)L_d^{1}$ is recorded. In particular, the loss changes in the model training process of four different weight $\lambda$ settings are mainly concerned. The forward results on the data of each batch along with the number of iterations are plotted below as Figure \ref{fig:loss_study}.

The change of the blue curve (i.e. \textit{Loss1}) are concerned in Figure \ref{fig:loss_study} because it refers to the main task. For all of the four settings, it drops rapidly after the training starts and reaches a lower level at around 300-th iterations, and there is no oscillation during the subsequent training process, which verifies the convergence of our model on the main task regardless of the weight $\lambda$ of auxiliary tasks. Another common phenomenon for all settings is that, in the process of \textit{Loss1} rapid decline, the curves of \textit{Loss2} and \textit{Loss3} showed an upward trend, so TotalLoss also showed an upward trend. It is believed that this is caused by the inconsistency of the objectives of different tasks in the multi-task joint training process, which is a normal phenomenon in multi-task learning. After the convergence of \textit{Loss1}, \textit{Loss2} and \textit{Loss3} begin to show a trade-off trend, the two loss show opposite changes in about 800 iterations in \ref{fig:loss_study} (a). After \textit{Loss3} reaches convergence, \textit{Loss2} begins to decline steadily, and the convergence process of TotalLoss is consistent with the convergence process of \textit{Loss2}. Similar trends can also be found in Figure \ref{fig:loss_study} (b), (c), (d). Compared with the other three $\lambda$ settings, the default setting ($\lambda_1=\lambda_2=\lambda_3=1$) achieves the best performance on the AUC metrics (0.78343), which verifies the rationality of our design of the loss function. By observing the results of different $\lambda$ settings, it can be concluded that reducing the weight $\lambda$ of \textit{Loss2} or \textit{Loss3} in the total loss lead to a certain degree of performance degradation of the model.

In summary, it is considered that the multi-task joint training of the model is essentially an alternative optimization process of different loss functions. Specifically, optimizing \textit{Loss1} during training may increase TotalLoss in the process of reducing \textit{Loss1}; \textit{Loss3} is optimized after \textit{Loss1} convergence, and this process may lead to the increase of \textit{Loss2}; Finally, the process of \textit{Loss2} decreasing until convergence is consistent with TotalLoss.

\section{Conclusion and Future Work} \label{Conclusion}

In this paper, an innovative deep learning cross-domain CTR prediction model named DADIN is proposed. The DADIN method is proposed, using a domain adversarial approach to enhance the knowledge transfer from a source domain to a target domain. The joint distribution alignment is designed by introducing a skip-connection-based domain agnostic layer and domain confusion loss method. The method is superior to other optimal single-domain and cross-domain models, achieving state-of-the-art results. Conceptual experiments on artificial datasets and ablation studies on real datasets demonstrate the effectiveness of the components of the model. In future work, a more reasonable neural network structure will be designed to realize the dynamic distribution alignment of both domains that can automatically adapt to the input and training process.


\printcredits

\section*{Declaration of competing interest}

The authors declare that they have no known competing financial interests or personal relationships that could have appeared to influence the work reported in this paper.

\section*{Acknowledgement}

This work is supported by High Performance Computing Center at Eastern Institute for Advanced Study. This work is also under the aegis of the Fundamental Research Funds for Central University of Central South University No. 2022zyts0611 and the Key program of National Social Science Foundation of China under Grant 22ATJ008.

\bibliographystyle{cas-model2}

\bibliography{DADIN}



\end{document}